\documentclass[referee]{aa} % for a referee version
\usepackage{epsfig}

\def\ser{R^{1/n}}
\def\dev{R^{1/4}}
\def\finf{f_{\infty}}
\def\Re{R_{\rm e}}
\def\Rexp{R_{\rm exp}}
\def\Rdev{R_{1/4}}
\def\Rmin{R_{\rm min}}
\def\Rmax{R_{\rm max}}
\def\Iz{I_0}

\def\Ime{\langle I \rangle_{\rm e}}
\def\Iobs{I_{\rm obs}}
\def\Imod{I_{\rm mod}}
\def\SBe{SB_{\rm e}}
\def\Lobs{L_{\rm obs}}
\def\Lmod{L_{\rm mod}}
\def\Lexp{L_{\rm exp}}
\def\Ltot{L_{\rm tot}}
\def\Ldev{L_{1/4}}
\def\Dmu{\Delta\mu}
\def\Dmumax{(\Dmu)_{\rm max}}
\def\Dmumean{\langle \Dmu \rangle}
\def\muobs{\mu_{\rm obs}}
\def\mumod{\mu_{\rm mod}}
\def\mudev{\mu_{1/4}}
\def\mumax{\mu_{\rm max}}
\def\mumin{\mu_{\rm min}}
\def\MB{M_{\rm B}}

\def\sigz{\sigma_0}
\def\sigr{\sigma_{\rm r}}
\def\sigobs{\sigma_{\rm obs}}
\def\ml{\Upsilon_*}
\def\Kvir{K_{\rm V}}

\def\Npar{N_{\rm par}}
\def\phiFP{\phi_{\rm FP}}
\def\magarsecs{{\rm mag/arc\sec {}^2}}

\catcode`\@=11
\def\gsim{\ifmmode{\mathrel{\mathpalette\@versim>}}
    \else{$\mathrel{\mathpalette\@versim>}$}\fi}
\def\lsim{\ifmmode{\mathrel{\mathpalette\@versim<}}
    \else{$\mathrel{\mathpalette\@versim<}$}\fi}
\def\@versim#1#2{\lower 2.9truept \vbox{\baselineskip 0pt \lineskip
    0.5truept \ialign{$\m@th#1\hfil##\hfil$\crcr#2\crcr\sim\crcr}}}
\catcode`\@=12
\begin{document}
   \title{Weak homology of elliptical galaxies}

   \subtitle{}

   \author{G. Bertin\inst{1,2},
           L. Ciotti\inst{1,3},
           \and
           M. Del Principe\inst{1,4}}

   \offprints{G. Bertin}

   \institute{Scuola Normale Superiore, Piazza dei Cavalieri 7,
              I-56126 Pisa, Italy
         \and
              Universit\`a degli Studi di Milano, Dipartimento di Fisica,
              via Celoria 16, I-20133 Milano, Italy
%              \email{Giuseppe.Bertin@unimi.it}
         \and
              Osservatorio Astronomico di Bologna, via Ranzani 1,
              I-40127 Bologna, Italy
%              \email{ciotti@bo.astro.it}
         \and
              Universit\`a degli Studi dell'Aquila, Dipartimento di Fisica,
              Via Vetoio Localit\`a Coppito, 67100 L'Aquila, Italy
%              \email{milena@terri1.te.astro.it}
                    }

   \date{}

\abstract{Studies of the Fundamental Plane of early--type galaxies,
from small to intermediate redshifts, are generally carried out under
the guiding principle that the Fundamental Plane reflects the
existence of an underlying mass--luminosity relation for such
galaxies, in a scenario where galaxies are homologous systems in
dynamical equilibrium. In this paper we reexamine the question of
whether a systematic non--homology could be partly responsible for the
correlations that define the Fundamental Plane.  We start by studying
a small set of objects characterized by photometric profiles that have
been pointed out to deviate significantly from the standard $\dev$
law. For these objects we confirm that a generic $\ser$ law, with $n$
a free parameter, can provide superior fits (the best-fit value of $n$
can be lower than 2.5 or higher than 10), better than those that can
be obtained by a pure $\dev$ law, by an $\dev$+exponential model, and
by other dynamically justified self--consistent models.  Therefore,
strictly speaking, elliptical galaxies should not be considered
homologous dynamical systems. Still, a case for {\it weak homology},
useful for the interpretation of the Fundamental Plane, could be made
if the best-fit parameter $n$, as often reported, correlates with
galaxy luminosity $L$, {\it provided the underlying dynamical
structure also follows a systematic trend with luminosity}.  We
demonstrate that this statement may be true {\it even in the presence
of significant scatter in the correlation $n(L)$}. Preliminary
indications provided by a set of ``data points" associated with a
sample of 14 galaxies suggest that neither the strict homology nor the
constant stellar mass--to--light solution are a satisfactory
explanation of the observed Fundamental Plane. These conclusions await
further extensions and clarifications, because the class of
low--luminosity early--type galaxies, which contribute significantly
to the Fundamental Plane, falls outside the simple dynamical framework
considered here and because dynamical considerations should be
supplemented with other important constraints deriving from the
evolution of stellar populations.  }

\authorrunning{Bertin et al.}
   \maketitle

\section{Introduction}

In the three-dimensional space of observed parameters $\log\sigz$,
$\log\Re$, and $\SBe = -2.5\log\Ime$ (where $\Ime = L/2\pi\Re^2$),
early--type galaxies are approximately located on a plane, called
the Fundamental Plane (hereafter FP; Dressler et al. 1987;
Djorgovski \& Davis 1987; Bender, Burstein, \& Faber 1992),
represented by the best--fit relation

\begin{equation}
\log\Re =\alpha \log \sigz + \beta\SBe + \gamma~.
\end{equation}

\noindent The central velocity dispersion $\sigz$ is often defined as
the velocity dispersion referred to an aperture radius of $\Re/8$
(J{\o}rgensen, Franx, \& Kj{\ae}rgaard 1993); the quantities $\Re$ and
$L$ are the empirically determined half--luminosity (effective) radius
and total absolute luminosity.  The coefficients $\alpha$, $\beta$,
and $\gamma$ depend slightly on the photometric band considered. By
measuring $\Re$ in kpc, $\sigz$ in km s$^{-1}$, and $\SBe = 42.0521
-2.5\log (L/ 2\pi\Re^2)$ in mag/arc sec$^2$, where $L$ is expressed in
units of the solar Blue luminosity, reported values are $\alpha
=1.25\pm 0.1$, $\beta =0.32\pm 0.03$, $\gamma =-8.895$ (J{\o}rgensen,
Franx, \& Kj{\ae}rgaard 1993; Bender et al. 1998; note that these
coefficients, in particular this choice of $\gamma$, refer to a {\it
long} cosmological distance scale). The FP is characterized by a small
scatter in $\Re$, on the order of 15\%, remarkably independent of the
position on the FP itself.

Studies of the FP of early-type galaxies, from low to intermediate
redshifts (for this latter case, see van Dokkum \& Franx 1996; Kelson
et al. 1997; van Dokkum et al.  1998a,b; Bender et al. 1998; Pahre, de
Carvalho, \& Djorgovski 1998; Pahre, Djorgovski, \& de Carvalho 1998;
J{\o}rgensen et al.  1999; Treu et al. 1999; Kelson et al. 2000a,b,c;
Treu et al.  2001a,b), are generally carried out under the guiding
principle that the FP reflects the existence of an underlying
mass--luminosity relation for such galaxies (see Faber et al.  1987;
van Albada, Bertin, \& Stiavelli 1995), in a scenario where galaxies
are homologous systems in dynamical equilibrium.  According to this
simplified picture, galaxies are thought to be characterized by a
universal internal structure (density and pressure tensor
distributions), which differs from galaxy to galaxy only in terms of
the relevant scales. An empirical indication that this may be the case
is suggested by the fact that, on the large scale, the luminosity
profile of elliptical galaxies appears to be universal.

The empirical $\dev$ luminosity ``law'' (de Vaucouleurs 1948) has
long been recognized to fit the surface brightness profiles of
elliptical galaxies successfully, to the point that, in the
absence of other (spectroscopic) indicators, elliptical galaxies
are routinely identified by means of this characteristic
photometric signature. In this sense, it might be argued that on
the large scale elliptical galaxies are characterized by a
universal luminosity profile and thus they are also likely to be
dynamically homologous systems. In practice, if we exclude the
central regions of the galaxy, generally affected by seeing or
instrumental PSF and sometimes studied separately in view of the
possible presence of a massive black hole, the overall $\dev$ fit
is characterized by residuals typically of the order of $0.1 -
0.2~\magarsecs$ (e.g., see de Vaucouleurs \& Capaccioli 1979;
Capaccioli 1987; de Carvalho \& da Costa 1988; Capaccioli 1989;
Burkert 1993). These deviations from the $\dev$ law, although
small, are often larger than the typical observational errors
involved. Interestingly, they are systematic and appear to
correlate with the galaxy luminosity (Michard 1985; Schombert
1986; see additional comments in the following paragraph). In
general, the surface brightness profile in the external regions is
found to be flatter for more luminous galaxies.

A simple and popular way to model the systematic non--homology
noted in the surface brightness distribution of early--type
galaxies is to consider an $\ser$ luminosity profile, with $n$ a
free parameter, as a generalization of the $\dev$ law (Sersic
1968; see also Davies et al. 1988; Capaccioli 1989; Caon,
Capaccioli, \& D'Onofrio 1993, hereafter CCD93; Young \& Currie
1994; D'Onofrio, Capaccioli, \& Caon 1994, hereafter DCC94;
Prugniel \& Simien 1997, hereafter PS97; Wadadekar, Robbason, \&
Kembhavi 1999). The $\ser$ ``law" has been found to be a
statistically convenient description to address some issues
related to the FP of elliptical galaxies (see Graham et al. 1996;
Ciotti, Lanzoni, \& Renzini 1996; Graham \& Colless 1997; Ciotti
\& Lanzoni 1997; Graham 1998). In addition, it has been applied to
the description of the surface brightness profiles of galaxy
bulges (see Andredakis, Peletier, \& Balcells 1995; Courteau, De
Jong, \& Broeils 1996). On the theoretical side, it has been the
focus of several general investigations (see, e.g., Makino,
Akiyama, \& Sugimoto 1990; Ciotti 1991; Gerbal et al. 1997;
Andredakis 1998; Ciotti \& Bertin 1999). If applicable, this
description would bridge the gap between the classical $\dev$
profile and the exponential profile characteristic of disks. It
has soon become clear that the shape parameter $n$ of the $\ser$
law correlates with global quantities such as total luminosity and
effective radius (CCD93; PS97; see also Khosroshahi et al. 2000).
For example, on a sample of 52 E/S0 galaxies with $\MB <-17.3$,
CCD93 find the following relation:

\begin{equation}
\log n =0.28 + 0.52\log\Re
\end{equation}

\noindent with $rms =0.18$. Here $n$ is the best-fit value
obtained by fitting the luminosity profile along the major axis,
and $\Re$ is determined from the curves of growth, independently
of the fit procedure. CCD93 noted that the transition from normal
to bright ellipticals occurs at $n=4$, $\Re = 3$ kpc, while
low--luminosity ellipticals (Davies et al. 1983) are characterized
by $n \approx 1$. The above relation implies the existence of a
similar relation between $n$ and $\MB$, since $\Re$ and $\MB$
correlate with each other. For example, CCD93 and DCC94 found:

\begin{equation}
n=-19.082-1.211\MB\simeq -19.082+3.0275\log L.
\end{equation}

\noindent The average scatter around this relation turns out to be
$\langle\Delta n\rangle\sim 1.895$, but the effective scatter
increases significantly with $L$.

Some confusion has accompanied the application of the $\ser$
models to observed objects. For example, PS97 report the
correlations $n-\Re$ and $n-\MB$ for a broad sample of
ellipticals; but here, in contrast with CCD93, the correlations
are shown to saturate, for $\log \Re [{\rm kpc}]  > 0.8$ and $\MB
< -20$, to values of $n$ close to 4. The apparent discrepancy is
to be traced to the fitting procedure used. In fact, PS97 referred
to the curves of growth (see comments in Appendix B.1 below). As a
result, for a given sample of early--type galaxies, values of $n$
not larger than 6 have been claimed by PS97, based on the use of
curves of growth, while values of $n$ up to 15 have been found
(see CCD93; DCC94) based on direct profile fitting. Curiously, for
the low-luminosity objects, where the latter procedure sets values
of $n$ below 5, the two methods appear to be mutually consistent.

Another natural way to address systematic deviations from the
$\dev$ law is to imagine (e.g., see Saglia et al. 1997) that the
galaxy results from the superposition of an $\dev$ component
(presumably spheroidal) and an exponential component (presumably
flat). This latter procedure is appealing from the dynamical point
of view (S0s are often thought to be made like this), yet the
procedure is usually applied as a purely photometric method (with
no check on its dynamical foundation; attempts in this latter
direction have been made, see Scorza \& Bender 1995).

In view of constructing a physical rationale at the basis of the FP,
the existence of strict homology is not necessary (see Ciotti \&
Pellegrini 1992; CCD93; Renzini \& Ciotti 1993; Djorgovski 1995;
Capelato, de Carvalho \& Carlberg 1995; Hjorth \& Madsen 1995; Ciotti,
Lanzoni, \& Renzini 1996; PS97; Pahre \& Djorgovski 1997; and
references therein). In fact, many of the arguments that allow for a
physical interpretation of the observed scaling law would carry
through even in the case that we may call of {\it weak homology}. This
is a condition by which the structure and dynamics (density and
pressure tensor distributions) of early--type galaxies vary
systematically with galaxy luminosity. An indication that this may be
a realistic picture is indeed provided by the systematic trends noted
above in terms of $\ser$ fits. However, we should emphasize that,
based on the photometric properties alone of early--type galaxies, we
cannot draw conclusions about the FP.  For this latter goal we have to
make use of dynamical models, because we need to translate the trends
observed in the photometric profiles into trends of the relevant
coefficients entering the virial theorem. The models used must at
least be generally consistent with the available kinematical
properties of early-type galaxies, otherwise the consequences drawn
would be based on unjustified models. An interesting attempt at
combining photometric and kinematic information towards a
comprehensive interpretation of the FP has been made recently by
Gerhard et al.  (2001). However, their detailed dynamical
investigation of a sample of 21 galaxies is far from being conclusive,
because their sample is mostly biased in the direction of luminous
objects and because their kinematical measurements, although based on
the sophisticated technique of fitting line profiles, are admittedly
subject to significant model limitations and large uncertainties.

We should emphasize that observations already provide important
empirical evidence both against strict homology and against the
constant stellar mass--to--light ratio hypothesis. On the one hand,
the existence of a relation such as that given in Eq.~(3) and the fact
that trends in the dark matter content with luminosity are noted
readily argue against strict homology. On the other hand, the
existence of clear correlations involving the galaxy luminosity, such
as the color-luminosity relation, and the fact that the coefficients
that define the FP depend on the observed waveband suggest that
systematic changes in the underlying stellar populations are
involved. At the interpretation level, an explanation based on a pure
age-luminosity relation is not viable, because it is hard to reconcile
with the small and approximately uniform scatter of the FP at zero
redshift and because it turns out to be incompatible with the observed
properties of the FP at intermediate redshifts. In addition, it has
been shown that interpretations based solely on either a dark
matter-luminosity relation or a metallicity-luminosity relation or a
systematic change of the Initial Mass Function with luminosity are not
satisfactory because they do not survive a close scrutiny. In this
paper we will add further internal consistency to the arguments in
favor of a form of weak homology at the basis of the FP.

In Section 2 we carry out an {\it in--depth} analysis of 4 significant
cases of deviations from the $\dev$ law by analyzing the surface
brightness profiles of NGC 1379, NGC 4374, NGC 4458, and NGC 4552. We
will confirm that a generic $\ser$ law, with $n$ a free parameter, can
provide superior fits (the best-fit value of $n$ can be lower than 2.5
or higher than 10), better than those that can be obtained by a pure
$\dev$ law, by an $\dev$+exponential model, and by other dynamically
justified self--consistent models. Although it may be argued that four
galaxies are a sample of little statistical significance, the detailed
tests that we have performed on these otherwise normal objects and,
especially, {\it the systematic trends with luminosity noted in the
literature for large samples} (see Eq.~(3), based on a sample of 52
galaxies) provide convincing evidence that homology cannot be
considered a {\it strict} rule for early--type galaxies.

In Section 3 we focus on the main implications of these conclusions on
the problem of the tilt and the thickness of the FP. For a constant
(from galaxy to galaxy) stellar mass--to--light ratio a suitable
correlation $n(L)$ could be consistent with the observed FP. We
demonstrate that this statement may be true {\it even in the presence
of significant scatter in the correlation $n(L)$}, by means of a
general mapping procedure from the space of the intrinsic physical
model quantities to the space of observed quantities. We also show
that, in principle, in the absence of strong empirical constraints on
the correlation $n(L)$, the Fundamental Plane could be realized in an
infinite number of ways, two of which are indeed the constant stellar 
mass--to--light solution, and the strict $\dev$ homology solution
combined with a suitable mass--luminosity relation. Actually,
indications provided by a set of ``data points" associated with a
sample of 14 galaxies are shown to suggest that neither the strict
homology nor the constant stellar mass--to--light solution are a satisfactory
explanation of the observed Fundamental Plane.

In the concluding Section 4, we briefly address some interesting
consequences on the expected trends of the observed FP at intermediate
redshifts and outline some natural possible extensions of the work
presented here.

Background work is recorded in the Appendices. Appendix A
summarizes the properties of the models used in the fits presented
in Section 2. Appendix B describes the relevant fitting
procedures. Appendix C illustrates some properties of isotropic
dynamical models in relation to the definition of the central
velocity dispersion $\sigz$. Finally, Appendix D records the
calculation of the virial coefficient associated with the
dynamical models considered in this paper.

\section{Deviations from the $\dev$ law}

\subsection{Preliminaries}

Many studies have already addressed the issue of the universality
of the $\dev$ law (e.g., see Burkert 1993; CCD93; DCC94; PS97; and
references therein). In this paper we would like to focus on some
relatively round objects, for which a significant departure from
the $\dev$ law has been noted. These objects will be used as
benchmark prototypes on which detailed tests will be performed
about the quality of a number of photometric models, about various
effects associated with the adopted fitting procedure, and about
the role of the radial range over which the photometric fit is
made.

In general, photometric profiles are compared with the $\dev$ law,
or with other photometric models, only over a finite radial
interval $(\Rmin,\Rmax)$. The inner limit is often taken so as to
avoid a detailed discussion of the effects of seeing or
instrumental PSF (the effects of the PSF on the observed surface
brightness profile become negligible beyond a radial distance from
the galaxy center of 2-3 times the scale of the relevant FWHM;
e.g., see Bertin, Saglia, \& Stiavelli 1988); furthermore, the
innermost region of a galaxy may present features that are related
to the presence of a massive central black hole, which is
generally studied as a separate issue. The outer limit often
reflects the quality of the available data, since special care has
to be taken to properly subtract the contribution of the sky
surface brightness (special techniques make it possible to measure
reliable photometric profiles down to $\mu_B\sim 28-29~
\magarsecs$). Of course, a correct estimate of the sky background
is considerably more difficult in the case of galaxies with nearby
companions.

Over a rather small radial range, a very good fit can be obtained
rather easily and it is thus hard to discriminate among different
models. For example, within the radial range $0.1\Re\leq R\leq
1.5\Re$, the galaxies NGC 1379 and NGC 4374 are well fitted by the
$\dev$ law. In fact, based on the definition of average residual
\begin{equation}
\Dmumean={1\over\sqrt N}\left\{\sum_{i=1}^N
[\mudev(R_i)-\muobs(R_i)] ^2\right\} ^{1/2},
\end{equation}
where $N$ is the number of photometric data points $\muobs$ inside the
radial range of the fit, and of maximum residual
\begin{equation}
\Dmumax =\max _{\Rmin\leq R_i\leq\Rmax}
         \left|\mudev(R_i)-\muobs (R_i)\right|,
\end{equation}
the galaxies NGC 1379 and NGC 4374 are found to be characterized
by $\Dmumean ~=0.033~\magarsecs$, $\Dmumax =0.079~\magarsecs$ and
by $\Dmumean ~= 0.052~\magarsecs$, $\Dmumax =0.108~\magarsecs$
respectively (Burkert 1993). As we will see, these results change
significantly when the radial range of the fit is increased (but
see Makino, Akiyama, \& Sugimoto 1990).
%--------------------------------------------------------------------
\begin{figure}[htbp]
\parbox{1cm}{
\psfig{file=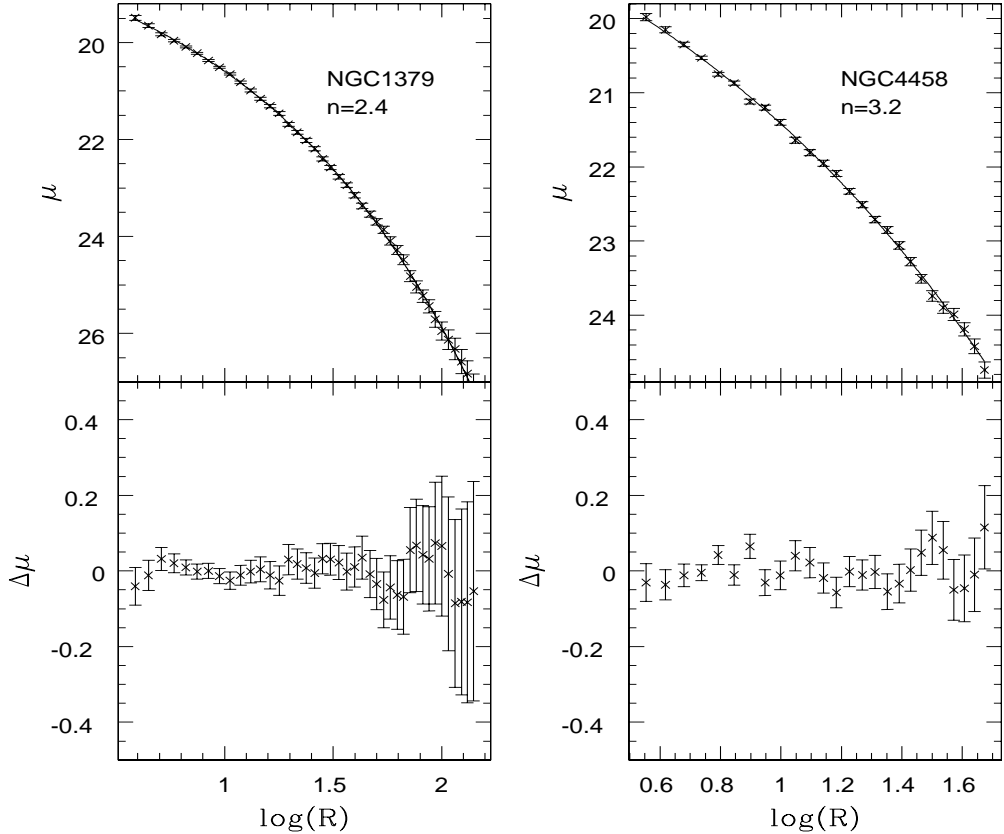,width=14cm,height=12cm,angle=0}}
\caption[]{$\ser$ fit. $R$ is in arcsec, $\mu$ in $\magarsecs$.  The
data points in Figs. 1-5 are in the Blue band, taken from Caon,
Capaccioli, \& Rampazzo (1990) and Caon, Capaccioli, \& D'Onofrio
(1994). Note the limiting surface brightness reached for each galaxy.}
\end{figure}
%-------------------------------------------------------------------

\subsection{The sample}

In the following we consider four apparently round ``prototypical
objects'', NGC 1379, NGC 4374, NGC 4458, and NGC 4552, that are known
to present significant deviations from the standard $\dev$ luminosity
distribution. We refer to photometric data (in the Blue band)
available in the literature, taken from Caon, Capaccioli, \& Rampazzo
(1990) and Caon, Capaccioli, \& D'Onofrio (1994), that have pointed
out the existence of a departure from the $\dev$ behavior. Indeed, we
have selected the four galaxies, from the sample of 80 galaxies
studied by these authors, based on the following criteria: (1) Round
isophotes; (2) Radially extended available photometry; (3) Significant
deviations from the $\dev$ law. At this stage, we are not interested
in the small--scale behavior of the luminosity distribution at very
small galactocentric radii and so we did not look up high--resolution
data from more modern investigations (e.g., from HST; see Stiavelli et
al. 2001), which would address issues beyond the goals of this paper.
On purpose, the adopted criteria 1)-3) do not refer to any, possibly
available, kinematical information on the four objects, because we
whish to conform to the usual (i.e., purely photometric) selection
rules adopted in the construction of early-type galaxy samples for
studies of the FP.

For this set of objects we consider the relative performance of four
different {\it photometric} models: (i) $\ser$ (one free dimensionless
parameter: $n$), (ii) $\dev$ (no free parameters), (iii)
$\dev$+exponential (two free parameters: $\Lexp/\Ltot$ and
$\Rexp/\Rdev$), and (iv) $\finf$ models (one free parameter:
$\Psi$). The main properties of the models are summarized in Appendix
A. In all cases, in addition to the best--fit values of the free
dimensionless parameters, if available, the fitting procedure leads to
the determination of two scales, i.e., of the {\it model} effective
radius $\Re$ and of the {\it model} total luminosity $L$.

Each of the four different photometric models is tested against a
variety of options for the fitting procedure: (a) profile fitting or
circularized profile fitting or curve of growth fitting, (b) reduced
or extended radial range, and (c) constrained or unconstrained
integrated luminosity. These fitting procedures are briefly described
in Appendix B.

\subsection{Relative performance of four photometric models for four
prototypical objects}

%--------------------------------------------------------------------
\begin{figure}[htbp]
\parbox{1cm}{
\psfig{file=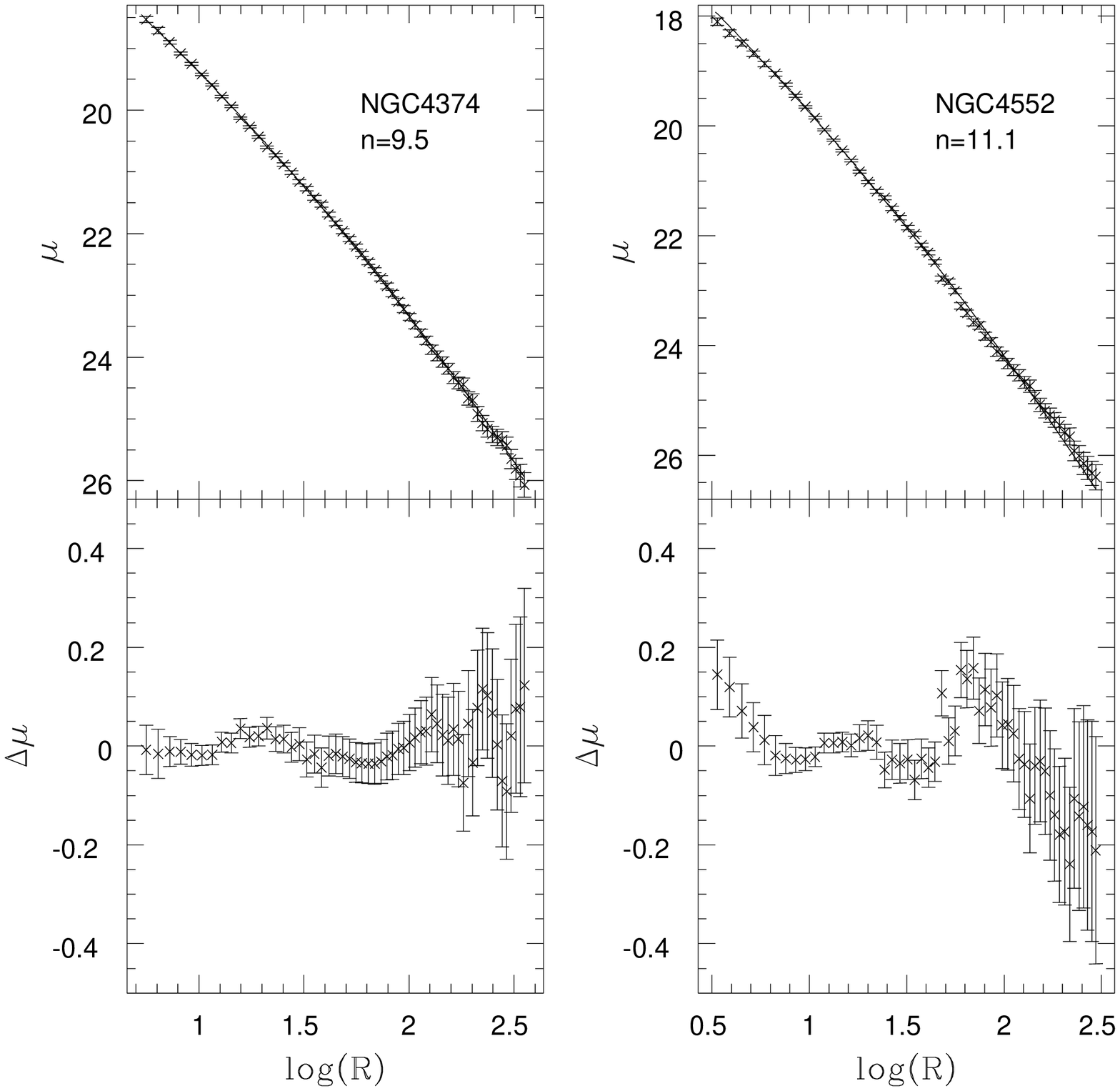,width=14cm,height=12cm,angle=0}}
\caption[]{$\ser$ fit. $R$ is in arcsec, $\mu$ in $\magarsecs$.}
\end{figure}
%-------------------------------------------------------------------

In Table 1 we record the results for the best-fit $\ser$ models,
for the reference option of using circularized profiles, of
considering the most extended radial range available, and of
imposing no constraint on the integrated luminosity. The quality
of the fit is measured by the parameter $\chi^2$, defined in
Appendix B, and is also described here by the values of $\Dmumean$
and $\Dmumax$ (see Eqs.~[4]-[5]). Some of the digits of the
various entries are not significant, but we have decided to keep
them since there should be no reason for misinterpretation. The
last two columns record the surface brightness excursion available
for the photometric data and the total absolute magnitude of each
object. The best-fit $\ser$ models are then illustrated in Figs. 1
and 2.

Similarly, in Table 2 we summarize the corresponding performance of
the other three photometric models (for the same option of fitting
procedure as in Table 1). It is clear that, especially for the two
objects (NGC 4374 and NGC 4552) that are characterized by high values
of $n$, the {\it model} effective radii thus determined by the fit can
vary dramatically (but see comments in Sect. 2.2). The fits are then
illustrated in Figs. 3-5. For the low-$n$ galaxy NGC 1379, the $\finf$
models are not providing a satisfactory fit, which is not surprising
since for this galaxy the role of rotation is likely to be important
(see the kinematical profile recorded by Graham et al. 1998) and thus
the galaxy falls outside the physical scenario at the basis of the
models. In addition, two minima in $\chi^2$ are actually available for
the $f_{\infty}$ models for this galaxy: the higher $\chi^2$ model
(lower $\Psi$; not illustrated in Fig. 5) is less bright than the
galaxy in the central region (opposite to the behavior of the model
fit shown in the Figure), both models are brighter than the galaxy at
large radii.

In Table 3 we summarize the results of the best-fit $\ser$ models
obtained under various fitting options. It is clear that many of
the quantities thus derived are sensitive to the adopted fitting
procedure. As an extreme example, we point to the case of NGC
4374, for which the best-fit model for the reduced radial range
(case $b$ of Table 3) would be characterized by $\Re = 1092$
arcsec, so that most of the total luminosity of the galaxy would
be assigned to the unobserved part of the photometric profile. Our
results quantify and generally confirm trends already noted in
earlier studies (see Introduction; but see also some important
remarks in Appendix B.1).

%___________________________________ Two column table (place early!)
\begin{table}[hbt!]
\caption[]{$\ser$ best fits. Circularized profiles, extended radial
range, no $L$ constraint. $\Re$ is expressed in arcsec.}
\begin{center}
\begin{tabular}{ccccccccc}
\hline
NGC  & $\chi^2$        & $\Re$                  & $\phiFP^{\mathrm{a}}$
     & $n$             & $\Dmumean$             & $\Dmumax$
     & $\mumax-\mumin$ & $M_B^{\mathrm{b}}$\\
\hline\hline 1379 & 0.31  & 25.7  & -5.33 & 2.38  & 0.042 & 0.086
& 7.64 & -19.51\\ 4374 & 0.46  & 259.0 & -5.08 & 9.48  & 0.044 &
0.123 & 7.54 & -21.35\\ 4458 & 0.85  & 22.2  & -5.57 & 3.19  &
0.044 & 0.115 & 4.76 & -18.45\\ 4552 & 1.32  & 104.0 & -5.13 &
11.14 & 0.095 & 0.239 & 8.30 & -20.73\\ \hline
\end{tabular}
\end{center}
\begin{list}{}{}
\item[$^{\mathrm{a}}$] $\phiFP$ is given by Eq. (6), with $\Re$ in arcsec and
$\SBe$ in $\magarsecs$.
\item[$^{\mathrm{b}}$] absolute magnitude in the Blue band (apparent magnitude
 and extinction from de Vaucouleurs et al.
(1991), adopted distance of 18.3 Mpc)
\end{list}
\label{Table 1a}
\end{table}
%___________________________________ Two column table (place early!)
\begin{table}[hbt!]
\caption[]{Comparison of best--fit models. Circularized profiles, extended
           radial range, no $L$ constraint. Columns from second to
           fourth refer to $\dev$ models, from fifth to ninth
           to $\dev$+exponential, and last four
           to $\finf$ models. $\Re$ is expressed in arcsec.}
\begin{center}
\begin{tabular}{cccc|ccccc|cccc}
\hline
NGC
& $\chi^2$
& $\Re$
& $\phiFP$
& $\chi^2$
& $\Re^{\mathrm{a}}$
& $\phiFP$
& $\Lexp/\Ltot$
& $\Rexp/\Rdev$
& $\chi^2$
& $\Re$
& $\phiFP$
& $\Psi$\\
\hline\hline
1379   & 5.85  & 34.9  & -5.33
       & 0.78  & 27.9  & -5.31 & 0.22 & 0.65
       & 3.22  & 32.6  & -5.30 & 9.6\\

       &       &       &
       &       &       &      &      &
       &(5.22) &(46.9) &(-5.38)&(5.2)\\

4374   & 9.05  & 75.2  & -4.96
       & 0.80  & 104.0 & -4.99 & 0.06 & 0.06
       & 1.56  & 98.5  & -4.94 & 7.3\\

4458   & 1.13  & 26.6  & -5.57
       & 0.74  & 21.4  & -5.55 & 0.13 & 0.96
       & 0.71  & 23.9  & -5.53 & 9\\

4552   & 12.23 & 41.2  & -5.07
       & 3.07  & 58.1  & -5.10 & 0.10 & 0.09
       & 2.86  & 146.4 & -5.22 & 6.3\\
\hline
\end{tabular}
\end{center}
\begin{list}{}{}
\item[$^{\mathrm{a}}$] effective radius of the combined profile
\end{list}
\label{Table2}
\end{table}
%___________________________________ Two column table (place early!)
\begin{table}[hbt!]
\caption[]{Best-fit $\ser$ models under various fitting options.
$\Re$ is expressed in arcsec.}
\begin{center}
\begin{tabular}{cccc|cccc|cccc}
\hline
NGC
& $n^{\mathrm{a}}$
& $\Re$
& $\phiFP$
& $\chi^2$
& $n^{\mathrm{b}}$
& $\Re$
& $\phiFP$
& $\chi^2$
& $n^{\mathrm{c}}$
& $\Re$
& $\phiFP$\\
\hline\hline
1379   & 2.88  & 26.8  & -5.32
       & 0.38  & 2.36  & 25.4  & -5.33
       & 0.31  & 2.39  & 25.7  & -5.33\\

4374   & 5.73  & 124.1 & -5.00
       & 0.33  & 14.5$^{\mathrm{d}}$  & 1092 & -5.25
       & 0.45  & 9.31  & 247.0 & -5.07\\

4458   & 4.61  & 29.8  & -5.57
       & 0.83  & 3.28  & 22.7  & -5.57
       & 0.81  & 3.21  & 22.3  & -5.57\\

4552   & 5.87  & 56.7  & -5.08
       & 0.85  & 9.95  & 95.2  & -5.12
       & 1.33  & 11.9  & 119.9 & -5.15\\
\hline
\end{tabular}
\end{center}
\begin{list}{}{}
\item[$^{\mathrm{a}}$]extended radial range, curves of growth, no
                      $L$ constraint (constant photometric error assumed, 
                      see App. B.1)
\item[$^{\mathrm{b}}$]reduced radial range ($\mumax-\mumin=4.5$ mag),
                      circularized profile, no $L$ constraint
\item[$^{\mathrm{c}}$]extended radial range, constrained luminosity $L$,
                      circularized profile
\item[$^{\mathrm{d}}$]solution ``crossing" (see description in App.~B.3)
\end{list}
\label{Table 1b}
\end{table}

For the set of four galaxies, considered to be as prototypes of
objects with a genuine deviation from the $\dev$ law, we have found
the following results. These are based on direct inspection of
Figs. 1-5 and especially on comparison of the $\chi^2$ values relative
to the various fitting options in Table 1 and 2.  For all galaxies the
fit by a pure $\dev$ law is the worst, with significant residuals. For
one galaxy (NGC 4458) the $\dev$+exponential model gives a marginally
better fit with respect to the $\ser$ model ($\chi^2=0.74$
vs. $\chi^2=0.85$), but for this galaxy also the pure $\dev$ law and
the $\finf$ models give reasonable fits ($\chi^2=1.13$ and 0.71,
respectively); not surprisingly, this is the galaxy with the smallest
surface brightness excursion $(\mumax -\mumin)$ sampled by the
available data (less than 5 magnitudes) and the one for which the
best--fit value of $n$ ($n \approx 3.2$) is closest to the standard
value of 4. For the three remaining galaxies (NGC 1379, NGC 4374, NGC
4552), for which the surface brightness profiles extend over a range
of more than 7 magnitudes, the $\ser$ model definitely gives the best
fit, even if it has fewer parameters than the $\dev$+exponential
model. In addition, the two objects (NGC 4374 and NGC 4552) that turn
out to be characterized by high values of $n$ (of order 10), when
fitted by the $\dev$+exponential model, would lead to a disk component
with an unusually high central brightness ($18.2$ and $18.1$
mag/arcsec$^2$ in the Blue band, respectively), at variance with the
central brightness of galaxy disks that characterize spiral galaxies.
It would be interesting to check wheter this is a common feature of
$\dev$ + exponential decompositions of early--type galaxies.

\subsection{Robustness of a photometric indicator}

In the discussion of the properties of the FP, it has often been
noted (see Kelson et al. 2000a; Treu et al. 2001a; see also
Trujillo, Graham, \& Caon 2001) that much of our uncertainties in
relation to the parameters derived by fitting the photometric
profile is resolved by an unexpected circumstance, that is the
empirical robustness of the photometric indicator that enters the
FP with respect to the modeling procedure. The relevant indicator
is

\begin{equation}
\phiFP =\log\Re - \beta\SBe~,
\end{equation}

\noindent with $\beta\approx 0.32$ (see Eq.~(1) and following
remarks).

An important point should be noted here. There are two different
ways of considering possible variations of $\phiFP$ for a given
object, on the basis of the definition of $\SBe$ reported just
before Eq.~(1). One is that of considering both $\Re$ and $L$ as
{\it model quantities}, both determined by the fit to the data
(see the comments on the extreme case of NGC 4374 made at the end
of the previous subsection). The other is to keep the total
luminosity as fixed (for the set of four galaxies studied earlier
in this paper, see the last column of Table 1), and let only $\Re$
be determined by the fit. In this latter case, it is readily shown
that the change in the value of the photometric indicator
associated with a small change of $\Re$ due to the modeling
procedure is $\delta \phiFP = (1 -5 \beta) \log{e} ~\delta \Re/\Re
\approx - 0.26~\delta \Re/\Re$; this shows that the change in
$\phiFP$ is, in this case, systematically smaller than the
relative change in $\Re$. No matter which definition of $\delta
\phiFP$ we refer to, the impact of a variation in $\phiFP$ on the
FP itself can be quantified by stating that, if we call ${\cal R}$
the ratio between the value of $\Re$ predicted by the FP, at a
given value of $\sigma_0$, and the observed value, then $\delta
{\cal R}/{\cal R} = - \delta \phiFP/ \log{e} \approx - 2.3~\delta
\phiFP$.

Here, based on the first definition of $\SBe$ (for which $L$ is
taken as a {\it model quantity}), we confirm the general validity
of the statement. Surprisingly, the photometric indicator is
robust both with respect to changes of the adopted photometric
model (compare Table 1 and Table 2) and with respect to variations
in the fitting procedure within the same photometric model
(compare Table 1 and Table 3). Therefore, for the purpose of {\it
measuring} the quantities entering the FP we are on safe grounds,
whereas for the purpose of {\it interpreting} the FP we are left
in a rather ambiguous situation.
%--------------------------------------------------------------------
\begin{figure}[htbp]
\parbox{1cm}{
\psfig{file=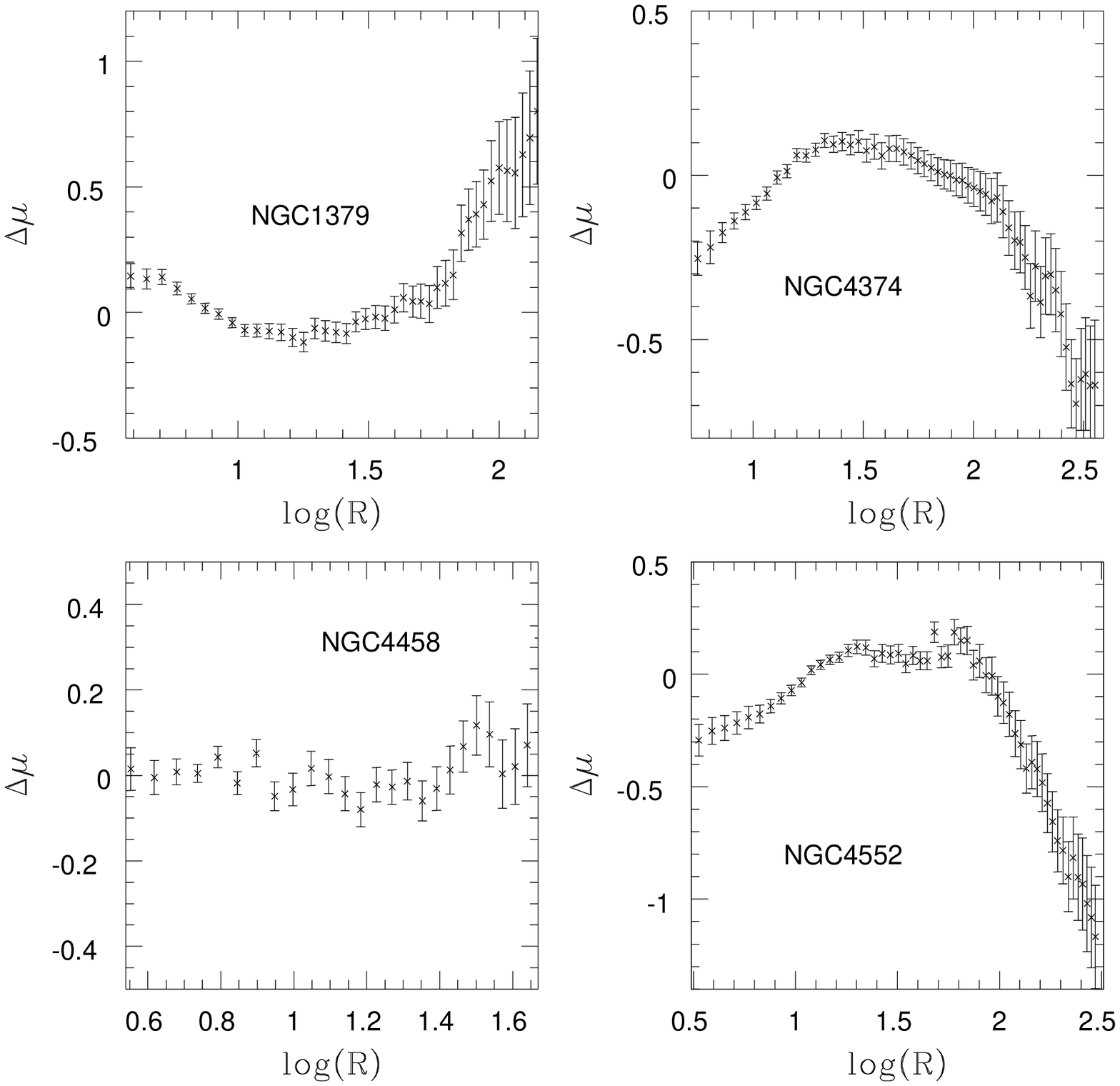,width=14cm,height=12cm,angle=0}}
\caption[]{$\dev$ fit. $R$ is in arcsec, $\mu$ in $\magarsecs$.}
\end{figure}
%-------------------------------------------------------------------

\subsection{The kinematical counterpart. Low-luminosity galaxies as
a physically separate class of objects}

The investigation provided earlier in this paper sheds some light
on the issues of modeling and confirms that homology cannot be
considered a strict rule for early--type galaxies.  However, the
discussion so far has been restricted to the photometric aspects,
while the possibility of {\it weak homology} (i.e., of systematic
trends in structural and dynamical properties as a function, e.g.,
of galaxy luminosity) should be assessed by considering the entire
set of quantities that enter the FP. In particular, it is
unavoidable that one should address the internal kinematical
properties of the galaxies involved. In fact, it has long been
known that low--luminosity ellipticals are likely to be rotation
supported systems (Davies et al. 1983). If we are interested in
drawing a connection between the low values of $n$ that best
describe low--luminosity ellipticals and the dynamical coefficient
that enters the virial theorem, we should not forget that the
virial theorem relation reflects the entire dynamical structure of
the system involved.

To be more specific, it is often argued that the relevant
dynamical coefficient can be conveniently calculated under the
simplifying assumption that the system be characterized by an
$\ser$ profile in projection, while intrinsically the system be
spherical, non-rotating, and strictly isotropic. From the Jeans
equations such coefficient is determined in a straightforward
manner (see PS97; Ciotti \& Lanzoni 1997; see also Sect. 3 and
Appendix D below). In practice, it remains quite doubtful that
such coefficient can be usefully inserted into a {\it realistic}
discussion of the FP, given the fact that the velocity dispersion
profiles thus calculated are often qualitatively different from
those observed (see Appendix C). Indeed, some of the difficulties
that we may find in placing the four galaxies described in this
section into the context of the FP, as will be attempted in
Sect.~3 (in particular, see Figs.~6-9), may just be due to this
basic modeling problem.
%--------------------------------------------------------------------
\begin{figure}[htbp]
\parbox{1cm}{
\psfig{file=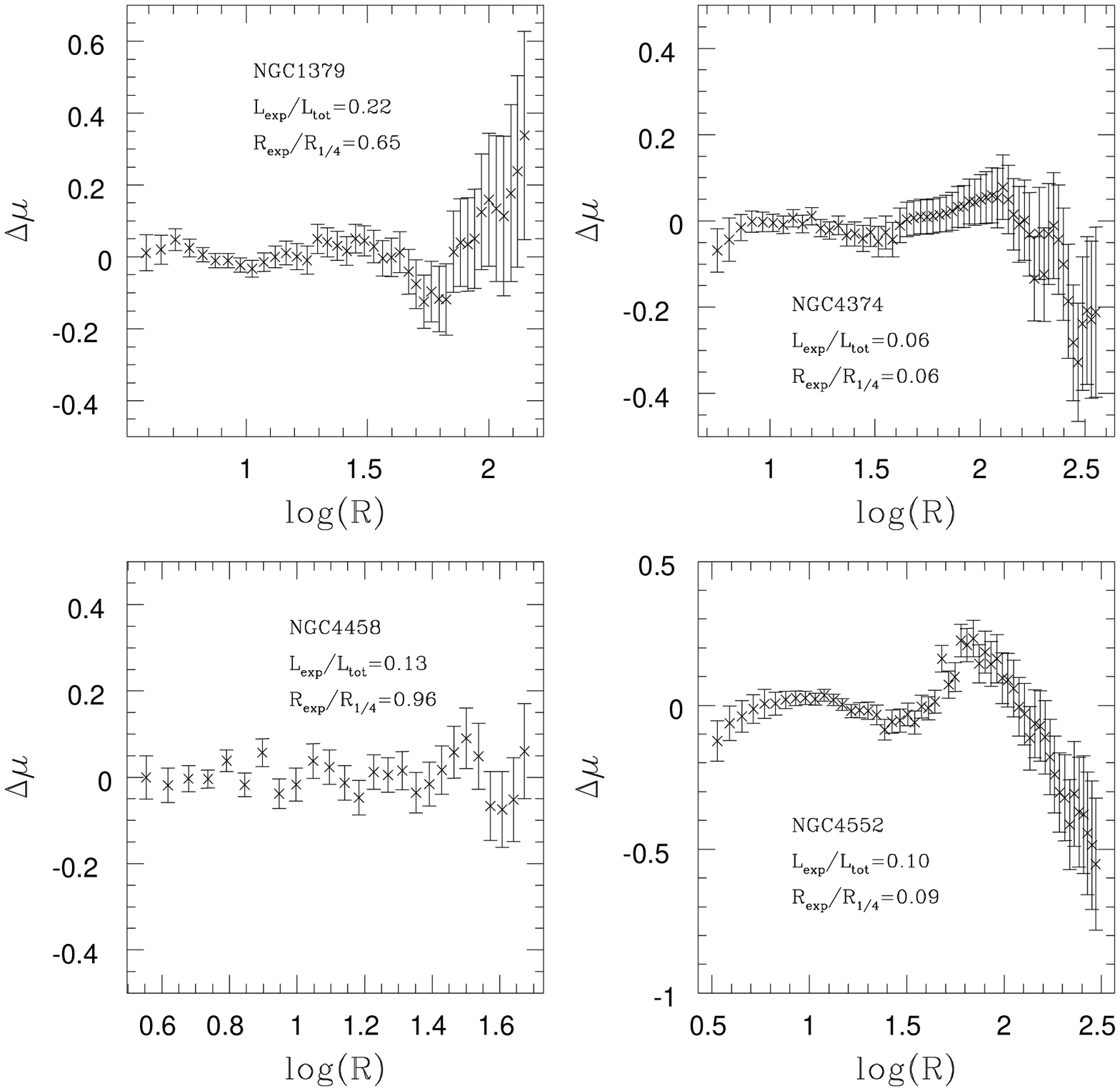,width=14cm,height=12cm,angle=0}}
\caption[]{$\dev$+exponential fit. $R$ is in arcsec, $\mu$ in
$\magarsecs$.}
\end{figure}
%-------------------------------------------------------------------

So far we may claim that we have a reasonable representation of
the dynamical structure only of bright ellipticals, at least in
the case of intrinsically fairly round objects (e.g., see Bertin
\& Stiavelli 1993). Low--luminosity ellipticals appear to be a
physically separate class of objects. Until we gain, by systematic
spectro--photometric surveys and by new and accurate modeling, a
good picture of how early--type galaxies are structured, from
faint to bright objects, there is no hope that we can make a
reliable discussion of the impact of the observed deviations from
the $\dev$ law on the FP. Nonetheless, in the absence of such
comprehensive dynamical modeling, by using the one--component,
spherical, non--rotating, isotropic $\ser$ models (even if we know
them to be an inadequate representation) we will be able to
demonstrate, in the next Section, that the absence of strict
$R^{1/4}$ homology can be consistent with an entire spectrum of
possible interpretations of the FP. Curiously, these
interpretations may well hold even if the empirical relation
$n=n(L)$ is affected by large scatter.

\section{Impact on the Fundamental Plane}

In order to assess the consequences of weak homology on the
interpretation of the FP, below we follow a line of thought
already introduced, e.g., by Ciotti \& Pellegrini (1992).

For a stationary stellar system, the scalar virial theorem can be
written as

\begin{equation}
{G\ml L\over\Re}=\Kvir\sigz^2~,
\end{equation}

\noindent where the coefficient $\Kvir$ takes into account
projection effects, the specific star orbit distribution and the
specific density distribution, and effects related to the presence
of dark matter; here $\ml$ is the {\it stellar} mass--to--light
ratio in the given band used for the determination of $L$ and
$\Re$. Equations (1) and (7) imply that in real galaxies, no
matter how complex their structure is, the dimensionless factor
$\ml/\Kvir$ is a well-defined function of any two of the three
observables $(L,\Re,\sigz)$. For example, by eliminating $\sigz$
from Eqs.~(1) and (7) we obtain

\begin{equation}
{\ml\over\Kvir}\propto \Re^{(2-10\beta+\alpha)/\alpha}
                        L^{(5\beta-\alpha)/\alpha}~.
\end{equation}

\noindent This fact is remarkable by itself.  All stellar systems
described by Eq.~(7) are in virial equilibrium, but only those
stellar systems for which $\ml/\Kvir$ scales according to Eq.~(8)
are placed onto the observational manifold given by Eq.~(1).  Of
course, this requirement is only necessary for a model to describe
real galaxies. In fact, the FP is not uniformly populated by
galaxies; instead, observed stellar systems populate only a
well-defined region of the FP.
%--------------------------------------------------------------------
\begin{figure}[htbp]
\parbox{1cm}{
\psfig{file=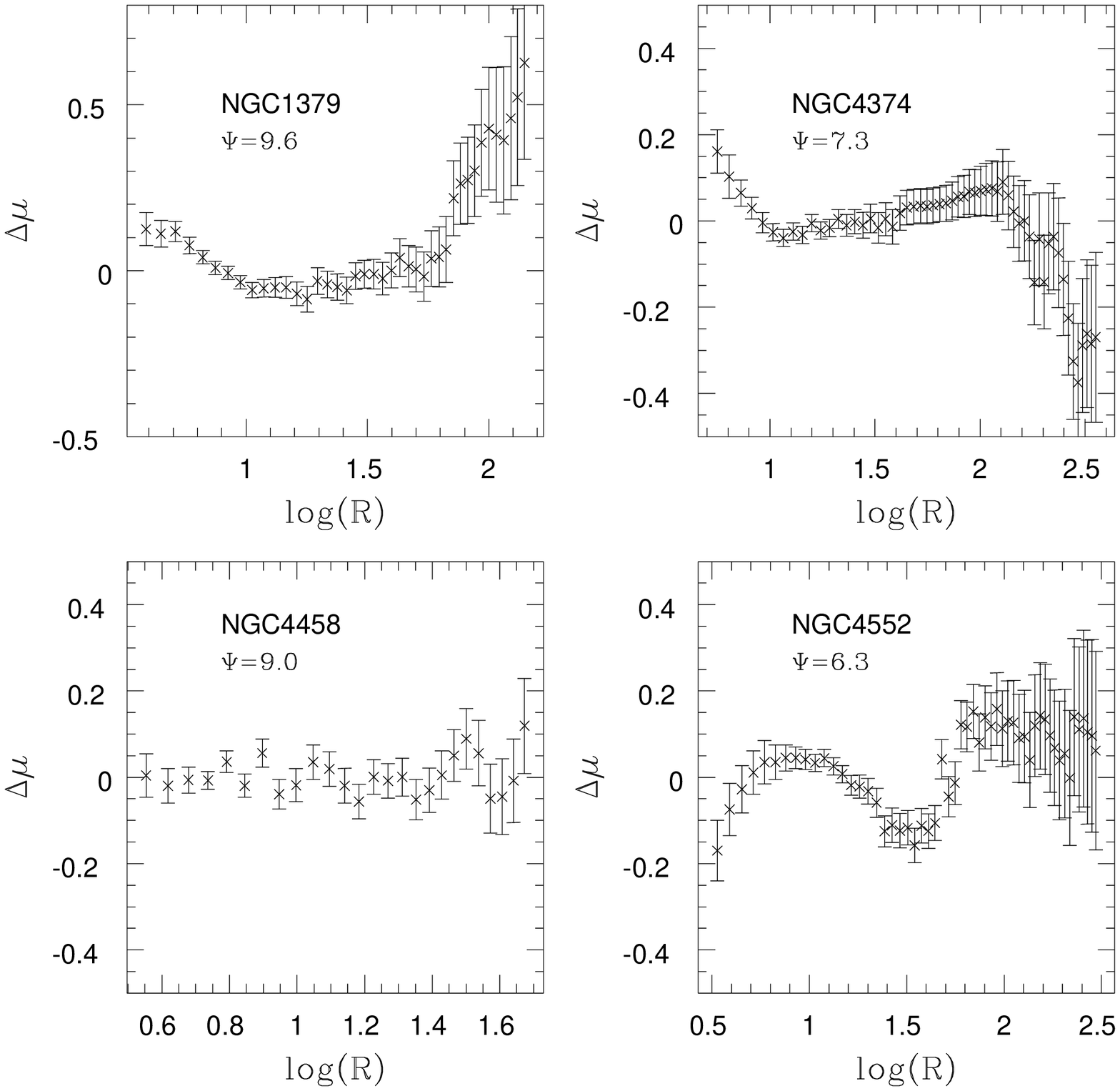,width=14cm,height=12cm,angle=0}}
\caption[]{$\finf$ fit. $R$ is in arcsec, $\mu$ in $\magarsecs$.}
\end{figure}
%-------------------------------------------------------------------

After recalling two simple extreme interpretations of the FP
(Subsection 3.1), in Subsection 3.2 we move to a more general
exploration, which will bring out once more a problem of fine
tuning met in the study of galaxies.

\subsection{Two extreme interpretations of the Fundamental Plane}

The possibility of simple extreme explanations of the FP is
associated with the fact that the exponent of the effective radius
appearing in Eq.~(8) is {\it observationally} very small, i.e.,
$(2-10\beta+\alpha)/\alpha\approx 0.04$ in the Blue band.

A first possibility to reproduce the scaling relation in Eq.~(8)
is then to assume homology (i.e., $\Kvir$ identical for all
galaxies), a variable stellar mass--to--light ratio (see, e.g., Renzini \&
Ciotti 1993; van Albada, Bertin, \& Stiavelli 1995), and the
empirically suggested identity $2-10\beta+\alpha=0$. With these
assumptions we find

\begin{equation}
\ml\propto L^{\delta}~;
\end{equation}

\noindent here $\delta=(2-\alpha)/2\alpha\simeq 0.30\pm 0.064$, where
the scatter has been calculated by standard error propagation
techniques. From a physical point of view, in a simple (coeval and
single metallicity) monolithic formation, passive evolution scenario,
a mass--luminosity relation for galaxies, of the kind suggested by
Eq.~(9), can result if, with significant fine tuning, the basic
properties of the Initial Mass Function, which describes the origin of
stellar populations, properly correlate with galaxy mass.

An alternative extreme explanation can be proposed by assuming a
constant stellar mass--to--light ratio $\ml$ and the existence of {\it
weak homology} (see Ciotti \& Pellegrini 1992; CCD93; Renzini \&
Ciotti 1993; Djorgovski 1995; Hjorth \& Madsen 1995; Ciotti,
Lanzoni, \& Renzini 1996; PS97; and references therein). In this
case it is the quantity $\Kvir$ (instead of $\ml$) that is
required to be a well-defined function of $\Re$ and $L$.  If we
take $2-10\beta +\alpha = 0$, the required dependence is

\begin{equation}
\Kvir\propto L^{-\delta}~,
\end{equation}

\noindent with $\delta$ and its scatter the same as above. As
noted in the Introduction, this latter interpretation is
qualitatively encouraged by the observations. Quantitatively, we
can easily show that we are in the right ballpark. In fact, from
Eq.~(10) it is realized that a factor of 20 in $L$, would require
a change in $\Kvir$ by a factor of $\approx 2.45$. If we consider
one--component, spherical, non--rotating, isotropic $\ser$ models
and a simulated spectroscopic aperture of $\Re/8$, the relation
$\Kvir(n)$ can be computed (Ciotti \& Lanzoni 1997; PS97; see
Appendix D). An accurate and convenient analytical representation
in the range $1\leq n\leq 10$ (with typical errors on the order of
one percent) is given by

\begin{equation}
\Kvir(n)\simeq {73.32\over 10.465 + (n-0.94)^2}+0.954~.
\end{equation}

\noindent It is then demonstrated that the observed excursion in
$n$ (see Sect.~2) could well be consistent with the desired range
in $\Kvir$. A similar argument also holds for the $f_{\infty}$
models, for which we have the following simple analytical
interpolation formula, accurate to around one percent,

\begin{equation}
\Kvir(\Psi)\simeq \frac{142.3 -41.51 \Psi + 2.66 \Psi^2}{30.61 -
10.7 \Psi + \Psi^2}
\end{equation}

\noindent in the range $2 \leq \Psi \leq 10$.

From this simple discussion we might conclude that weak homology of
elliptical galaxies might indeed be consistent with a constant $\ml$
and the FP.

Unfortunately, the determination of the coefficient $\Kvir$ is
sensitive to the choice of models (see the expressions recorded
above and App.~D) and to the choice of the aperture radius. For
the $\ser$ models, the latter dependence is not too significant,
even in the presence of the well-known unrealistic central
depression in their projected velocity dispersion profiles (see
App.~C); this has been checked by comparing the values of $\Kvir$
obtained on the basis of an $\Re/10$ aperture with those based on
$\Re/8$. In order to make a physically better justified case, one
might think it more appropriate to refer to the $\finf$ models,
which are known to possess realistic kinematical profiles. Note
that this would lead to significantly different answers with
respect to the application of $\ser$ models. For example (see
App.~D), if we take NGC 4552, the $\ser$ modeling would give
$n\simeq 11.14$ and $\Kvir\simeq 1.6$, while the $\finf$ modeling
would set $\Psi\simeq 6.2$ and $\Kvir\simeq 2.5$; for the galaxy
NGC 4458 (closest to the standard $\dev$ law in our sample), we
find $n\simeq 3.2$ and $\Kvir\simeq 5.7$ based on $\ser$ modeling
or $\Psi\simeq 9$, $\Kvir\simeq 3$ based on $\finf$ modeling. Here
we prefer not to pursue this line of investigation, because the
empirical relation $\Psi=\Psi (L)$ has not yet been studied
systematically and because, in any case, we know a priori that
such modeling, even if it might turn out to be partially
successful, would lack physical justification when applied to
low--luminosity, rotation--supported galaxies.

\subsection{A more general interpretation allowed by weak homology}

We now describe a more general interpretation of the FP allowed by
the existence of weak homology. For simplicity, we refer here to
the family of $\ser$ models that has led to Eq.~(11), although the
method proposed could be implemented for any well--defined family
of dynamical models.  In particular, here we explore the
theoretical manifold obtained by mapping, with the aid of Monte
Carlo simulations, the model space $(n,\ml,L,\Re)$ into the
observed space $(L,\Re,\sigz)$. The domain of model galaxies
considered in the simulations is defined by $1~\lsim~
\Re~\lsim~20$, $0.1~\lsim~L~\lsim~50$, and
$100~\lsim~\sigz~\lsim~35   0$, where radii are measured in kpc,
luminosities in $10^{10}$ solar Blue luminosities, and velocities
in km s$^{-1}$. For each four-dimensional point $(n,\ml,L,\Re)$ in
model space we calculate $\sigz$ from Eq.~(7) and then we check if
the mapped (virialized) system belongs to the FP. This allows us
to select in model space, as acceptable candidates for real
galaxies, only those points that turn out to be compatible, in the
observed space, with the FP correlation and the observed scatter
around Eq.~(1). We also explore the ``backbone" model manifold
obtained by artificially reducing the FP thickness to a limiting
zero--scatter case. For simplicity, we plot the projections of the
manifold thus derived on the natural planes $(L, \ml)$ and $(L,
n)$, which are precisely the two planes that are used to
illustrate the two extreme interpretations described in Subsection
3.1.

We start this exploration by applying the proposed technique to
the two cases discussed in the previous subsection.  The case of
$R^{1/4}$ homology is studied by restricting the model population
to a set of points for which $3.99\leq n\leq 4.01$. The derived
dependence of $\ml$ on $L$ is well fitted by the power law
relation of Eq.~(9), with $\delta \simeq 0.27$ (consistent with
the value $0.30 \pm 0.064$ argued in Subsection 3.1), with reduced
$\chi^2\simeq 1.7$. The scatter of points along the vertical axis
at fixed $L$ (see Fig.~6) reflects the assumed {\it thickness} of
the FP. Indeed, if we artificially reduce the scatter in $\Re$
from 0.15 to 0.01, the derived power-law exponent becomes $\delta
= 0.28$, with reduced $\chi^2\simeq 7\times 10^{-2}$ (the thick
locus of points in Fig.~6).

%--------------------------------------------------------------------
\begin{figure}[htbp]
\parbox{1cm}{
\psfig{file=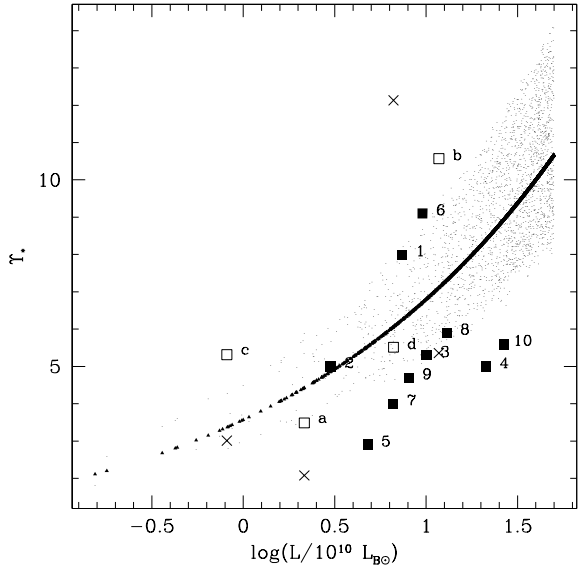,width=8cm,height=8cm,angle=0}}
\caption[]{Projection of the manifold in model space identified
           by $\dev$ homology. The scatter of points reflects
           the adopted scatter around the FP in the observed space.
           The symbols that trace the thick line represent the ``backbone"
           manifold. Open squares represent the four galaxies of Sect.~2,
           following the $\ser$ modeling summarized in Tables 1
           and D1; crosses represent the same four galaxies,
           following the $\finf$ modeling summarized in
           Tables 2 and D1 (see Subsect.~3.3 for additional description).
           Filled squares represent the sample of ten galaxies considered
           in Table 5 of Bertin et al. (1994); see Subsect.~3.4 of the present
           paper for a description.}
\end{figure}
%-------------------------------------------------------------------

To illustrate the second case (see Fig.~7), we pick a constant stellar
mass--to--light ratio $3.49 \leq \ml \leq 3.51$. The selected points
are quite dispersed along the $n$ axis. A good fit is obtained by
a law of the form $n = a + b \log{L}$, with $b\simeq 4.02$ and
$\chi^2\simeq 1.3$, not so well in agreement with the observed
scaling relation given in Eq.~(3). After artificially reducing the
scatter around the FP to 1 percent in $\Re$, we now have $b \simeq
5.03$ and $\chi^2\simeq 0.24$. (By increasing or decreasing the
reference value of $\ml$, the simulation points would move below
or above the strip identified by the present choice of $\ml$.)
%--------------------------------------------------------------------
\begin{figure}[htbp]
\parbox{1cm}{
\psfig{file=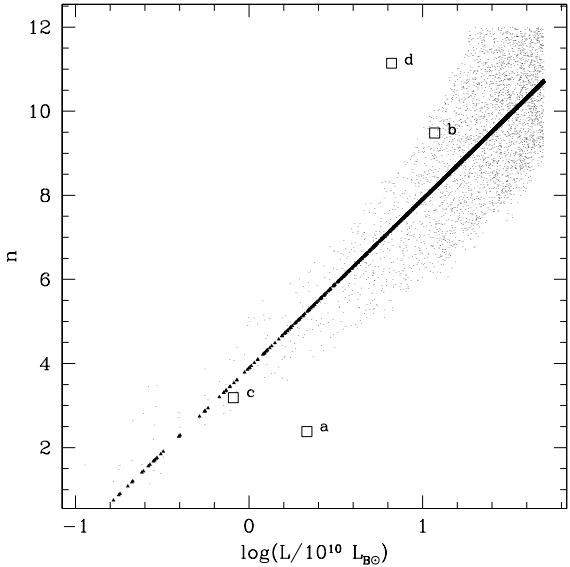,width=8cm,height=8cm,angle=0}}
\caption[]{Projection on the plane $(L,n)$ of the manifold in
           model space identified by the constant mass--to--light
           ratio assumption $\ml\simeq 3.5$. The scatter of points
           reflects the adopted scatter around the FP in the observed space.
           Solid symbols represent the ``backbone" manifold.
           Open squares represent the four galaxies of Sect.~2,
           following the $\ser$ modeling summarized in Tables 1
           and D1 (see Subsect.~3.3 for additional description).}
\end{figure}
%-------------------------------------------------------------------

A somewhat unexpected result of these simple simulations is that
the relatively small dispersion of $\Re$ for real galaxies, around
the value associated with the FP at any fixed $L$ and $\sigz$, is
consistent with a significantly larger dispersion of the model
parameters along the vertical axes in Figs.~(6) and (7) at fixed
values of $L$. Therefore, weak homology (e.g., a correlation
between the structural parameter $n$ and the luminosity $L$) can
be at the origin of a relatively thin FP, even in the presence of
a substantial dispersion of $n$ at given $L$.

At this point we can proceed to consider the most general case,
where neither homology is strict nor mass--to--light ratio is
constant.  In fact, we leave the model points to run over the
entire ranges of $\ml$ values and $n$ values in the domain
indicated at the beginning of this subsection. For simplicity, we
look for the backbone manifold selected by artificially reducing
the scatter on the observed $\Re$ to 0.01. In this way the scatter
presented by the points on the two natural planes $(L, \ml)$ and
$(L, n)$ will be determined only by projection effects (of the
relevant manifold onto the desired planes).
%--------------------------------------------------------------------
\begin{figure}[htbp]
\parbox{1cm}{
\psfig{file=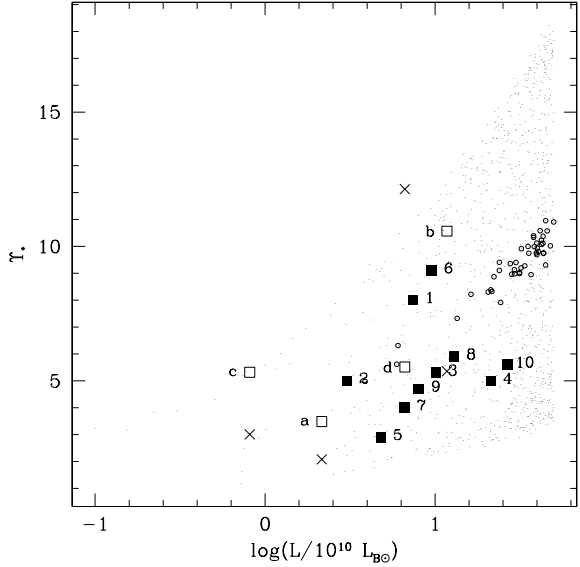,width=8cm,height=8cm,angle=0}}
\caption[]{Projection on the $(L,\ml)$ plane of the entire
           backbone manifold identified by the FP with scatter
           artificially reduced to 0.01. Open circles correspond
           to models with $3.8\leq n\leq 4.2$. Open squares, crosses, and
           filled squares define observed galaxies as in Fig.~6.}
\end{figure}
%-------------------------------------------------------------------

In Fig.~8 we show the projection of what we have called the
backbone manifold onto the $(L,\ml)$ plane. Note how, at any given
value of $L$, a larger range of $\ml$ values is permitted, even if
the FP considered in the simulations is extremely thin. Therefore,
in the general case, $\ml$ is not necessarily well correlated with
galaxy luminosity, as a result of the presence of the free
parameter $n$. It is interesting to check how models with
different $n$ are distributed inside this projection of the
backbone manifold; the open circles represent models characterized
by $3.8\leq n\leq 4.8$, a range slightly broader than that shown
in the homology case shown in Fig.~6. In any case, also under the
more general circumstances considered here, the FP relation is
associated with an overall increasing trend for the stellar mass--to--light
ratio $\ml$ with $L$.

%--------------------------------------------------------------------
\begin{figure}[htbp]
\parbox{1cm}{
\psfig{file=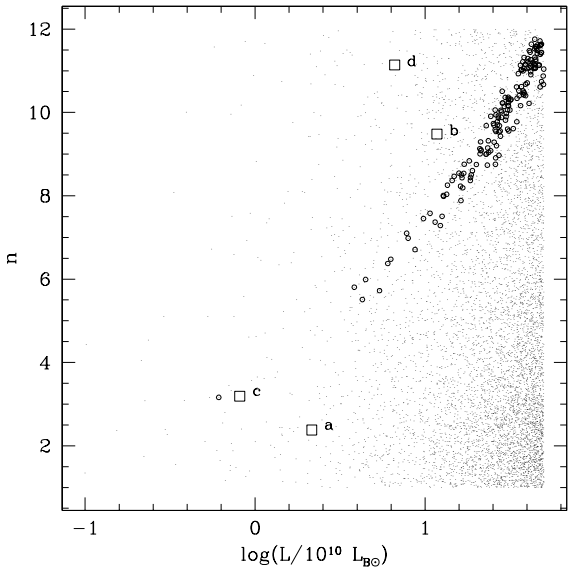,width=8cm,height=8cm,angle=0}}
\caption[]{Projection on the $(L, n)$ plane of the entire
           backbone manifold identified by the FP with scatter
           artificially reduced to 0.01. Open circles correspond
           to models with $3.4 \leq\ml\leq 3.6$. Open squares
           define observed galaxies as in Fig.~7.}
\end{figure}
%-------------------------------------------------------------------

Similar comments apply to the projection illustrated in Fig.~9.
Here the open circles represent models with the stellar mass--to--light ratio
$\ml \approx 3.5$. Note the large spread in $n$ present, still
consistent with the existence of a thin FP, once the restriction
of nearly constant $\ml$ is relaxed. Note also that there remains
only a mild trend of $n$ with $L$ required by the existence of the
FP, but only if we focus on the zero-scatter limit. Since in
reality some correlation is observed (see Introduction and
Eq.~(3)), we should conclude that something else, in addition to
the FP, characterizes the underlying weak homology of real
galaxies.

\subsection{Locating the four prototypical objects 
without detailed stellar dynamical modeling}

With the aim of a better appreciation of the naive indications
provided by a {\it pure photometric modeling} we have recorded, in
Figs.~6-9, as open squares (labels $a$, $b$, $c$, $d$ following the
order of Table 1), the points available for the set of four galaxies
discussed in the previous part of the paper and from the available
data about their central velocity dispersions. This information (see
Table D.1), based on the $\ser$ modeling, allows us to extract values
of $\ml$ from the observed values of $\sigz$ (119, 296, 106, and 269
km s$^{-1}$ for NGC 1379, 4374, 4458, and 4552, respectively, McElroy
1995).  In Fig.~6 and Fig.~8 we also record the different points that
would be argued on the basis of a {\it one-component} $\finf$
photometric modeling. 

For the purposes of this Section, the relevant quantities associated
with the four galaxies are rescaled, when necessary, to the long
cosmological distance scale. This is consistent with the adopted
choice of coefficients for the FP (see Eq.~(1) and following comments)
and with the analysis to be presented in the following Subsection.

\subsection{Trends from a sample of relatively bright 
galaxies and the role of dark matter}

In order to better relate the arguments provided in this Section
with the trends observed in real galaxies, we have included in
Figs.~6 and 8 some ``data-points" that are external to the
simulations described above. These data-points have been obtained
as follows.

We have referred back to the results of an ESO Key Programme ``A
search for dark matter in elliptical galaxies" and, in particular, to
a set of 10 galaxies the properties of which are summarized in Table 5
of Bertin et al. (1994). The objects are: NGC 1399, NGC 3379, NGC
4374, NGC 4472, NGC 5812, NGC 5813, NGC 7507, NGC 7626, NGC 7796, and
IC 4296 (some of these are also part of the sample considered by
Gerhard et al. 2001). The numbers from 1 to 10 of the filled squares
in Fig.~6 and Fig.~8 represent precisely these objects in the order
followed above. These galaxies had been considered as those for which
extensive and reliable {\it stellar dynamical} modeling was available,
on the basis of {\it accurate and radially extended} stellar dynamical
data. We borrow the values of $\ml$ and $\Re$ listed in Table 5 of
Bertin et al. (1994) under the columns $(M_{\rm L}/L_{\rm B})_{2C}$
and $\Re^m$, respectively, and thus obtain the values of $\Kvir$ for
those objects, based on a detailed {\it stellar dynamical modeling}.
One object, NGC 4374, is also part of the set of four galaxies for
which we have re-examined the {\it photometric} profile in Sect.~2.

We note that those models did include significant amounts of {\it
dark matter}. Therefore, it is not surprising to find now that
several filled squares fall outside the strip outlined in Fig.~6,
which has been calculated on the basis of strict $\dev$ homology.
Then Fig.~8 shows that the same data-points are well consistent
with a possible weak homology that is neither the strict $\dev$
homology nor the one based on a constant $\ml$. Note that the
broad area outlined in Fig.~8 has been obtained by reducing
artificially the scatter of the FP to 0.01.

\section{Conclusions}

This paper, organized in two parts, has led to a number of
results, although many are the questions that remain only partly
answered. Here we provide a summary of the main points made in the
paper and of the most important questions that are left for future
investigations.

In the first part of the paper we have re-examined the surface
brightness profiles of NGC 1379, NGC 4374, NGC 4458, and NGC 4552,
four galaxies that are known to present significant deviations
from the $\dev$ law. We have considered the relative performance
of four different photometric models, namely the $\ser$, $\dev$,
$\dev$+exponential, and $\finf$ models. Each of the four different
photometric models has been tested against a variety of options
for the fitting procedure. We have thus reached the conclusion
that there is indeed convincing evidence that homology cannot be
considered a strict rule for early type galaxies. In particular,
we have found that:

\begin{itemize}
  \item For the four galaxies of this limited sample the fit by a pure $\dev$ law is the
worst.
  \item For three objects the $\ser$ photometric model definitely provides
the best fit, even better than that by the $\dev$+exponential
model, although the latter model has the advantage of one
additional free parameter.
  \item For NGC 4458, which is the object with the smallest surface brightness excursion
available (less than five magnitudes), the $\dev$+exponential
model gives a marginally better fit with respect to the $\ser$
model; but, for this, the $\dev$ and the $\finf$ models give
practically equivalent fits.
  \item The $\finf$ models manage to
provide a reasonable description of the bright ellipticals even
when the $\ser$ modeling suggests a value of $n$ significantly
larger than 4. However, they do not appear to be viable for the
low-luminosity ellipticals, which are characterized by a small
value of $n$. This does not come as a surprise, because the
physical arguments that have led to the construction of the
$\finf$ models reflect the picture of low rotation, collisionless
collapse, while low-luminosity ellipticals are known to be
generally rotation supported.
  \item The values of the effective radius resulting from the fit can change
dramatically (by factors of up to three), depending on the fitting
procedure and on the adopted photometric model. On the other hand,
we confirm that the determination of the photometric indicator,
i.e. the combination of photometric quantities entering the FP, is
robust.
\end{itemize}

The good performance of the $\ser$ model in describing the {\it
photometric} profiles of galaxies deviating from the $\dev$ law
and the existence of an empirical correlation between $n$ and $L$
(see Eq.~(3), based on a sample of 52 galaxies) suggest that
elliptical galaxies might be considered as weakly homologous
systems.  Of course, a clear indication of weak homology cannot be
based on photometric properties alone. In the second part of the
paper, we have then addressed the impact of weak homology on the
physical interpretation of the FP, starting from a re-discussion
of the rationale behind two extreme interpretations of the tilt of
the FP. One is obtained by assuming strict homology and a suitable
relation between the stellar mass--to--light ratio and the galaxy
luminosity. The other is obtained by taking a constant
stellar mass--to--light ratio (from galaxy to galaxy) and a suitable form
of weak homology. The latter interpretation usually raises
concerns (besides the problems related to the interpretation of
the observed correlations involving luminosity, such as the
color-luminosity relation; see discussion in the Introduction),
because the large scatter around the correlation $n =n(L)$ appears
to be hard to reconcile with the small dispersion observed around
the FP. We have thus obtained the following results:

\begin{itemize}
  \item In order to address the problem in its generality, we have
developed a systematic approach to the study of the FP, based on a
mapping between model and observed quantities. For simplicity, we
have used, as a guiding tool for the present investigation, simple
one-component, spherically symmetric, isotropic $\ser$ models. The
study would be best carried out with other families of models,
better justified from the physical point of view, but for the
present purposes the simple $\ser$ models used provide an adequate
demonstration. The use of one-component models is equivalent to
assuming that dark matter, if present, is distributed in direct
proportion to the luminous matter.
  \item Within this framework, we have shown how
the scatter of the stellar mass--to--light ratio (in the first of the two
simplest interpretations) or of the index $n$ (in the second
interpretation) is fully controlled by the observed thickness of
the FP. This quantifies in detail the problem of parameter fine
tuning that has to be faced.
  \item We have then relaxed the hypothesis of a simple (i.e., one-parameter)
origin of the tilt of the FP and focused, for simplicity, on the
structure of the FP with artificially reduced thickness. A fine
tuning problem obviously remains, because the existence of the FP
imposes a strict condition on $\ml/\Kvir$. However, we have been
able to demonstrate that the ranges of variation for the stellar 
mass--to--light ratio and the index $n$ of the models, at fixed
galaxy luminosity, can be substantial and still be consistent with
an arbitrarily thin FP. The fine tuning involved requires that
variations in the stellar mass--to--light ratio be compensated for by
corresponding variations in galaxy structure.
  \item Preliminary indications provided by a set of ``data points"
associated with a sample of 14 galaxies suggest that neither the
strict homology nor the constant stellar mass--to--light solution are a
satisfactory explanation of the observed Fundamental Plane.
\end{itemize}

This paper opens the way to the following new investigations, to
be carried out in separate papers:

\begin{itemize}
 
\item A study based on the approach introduced in Sect.~3 within a
more complex dynamical structure (presence of dark matter and more
complicated geometries for the model galaxies).

\item A study of the evolution of the FP with redshift, within a
physical interpretation based on weak homology. If the observations
were able to tell the time evolution of the coefficients $\alpha$ and
$\beta$ in the exponents in Eq.~(8), then, from our mapping procedure,
we could infer the actual evolution shared between luminosity and
structural parameters.  From a different, predictive point of view, if
we argue that the tilt of the FP is totally due to weak homology (with
constant $\ml$ from galaxy to galaxy, so that the FP is generated by
systematic structural variations with luminosity of the quantity
$\Kvir$), then we would expect only a parallel shift of the FP,
i.e. only a systematic shift of its zero-point $\gamma$ with time.
Clearly, this picture must be {\it only approximately} viable, because
galaxies follow the metallicity - velocity dispersion relation (see
Bender, Burstein, \& Faber 1993 and references therein) and the
color-magnitude relation (see Bower, Lucey, \& Ellis 1992), consistent
with the fact that the tilt of the FP depends on waveband (e.g., see
the results based on near-infrared data obtained by Pahre, Djorgovski,
\& de Carvalho 1998). On the other hand, if, as is frequently done, we
refer to the picture where all galaxies are strictly homologous
systems, so that the tilt of the FP is completely due to a luminosity
dependence of $\ml$ (at fixed metallicity), then the luminosity
evolution of faint galaxies should be faster than that of bright
galaxies, and the tilt of the FP would be expected to change with $z$,
with the larger changes at the FP faint end. Here several different
physical scenarios would be available (see also Renzini \& Ciotti
1993).

\item A study of the relation between weak homology and the popular
paradigm of galaxy merging.

\end{itemize}

From the observational point of view, it is especially urgent to carry
out a discussion of weak homology, extended to the spectroscopic
(stellar-dynamical) data, for a complete sample of nearby early-type
galaxies.

\begin{acknowledgements}
We would like to thank T. van Albada, R. de Carvalho, M. Lombardi, and
T.  Treu for very useful comments and suggestions. This work has been
partially supported by MURST of Italy (co-fin 2000).
\end{acknowledgements}

\appendix

\section{Photometric Models}

\subsection{$\ser$ and $\dev$ models}

Consider the surface brightness profile (Sersic 1968) given by

\begin{equation}
I(R)=\Iz e^{-b\eta^{1/n}},
\end{equation}

\noindent where $\eta=R/\Re$, $n$ is a positive real number, and
$b = b(n)$ a dimensionless constant such that $\Re$ is the
effective radius, i.e., the projected radius encircling half of
the total luminosity associated with $I(R)$. As proved in a
separate paper (Ciotti \& Bertin 1999), the first terms of the
asymptotic expansion of $b(n)$, for large $n$, are $b(n)\sim
2n-1/3 +4/(405 n)$, already accurate to a part in one thousand for
$n$ as low as unity. The proper normalization of the R$^{1/n}$
profile, to be considered for a case with given scales $L$ and
$\Re$, is

\begin{equation}
I(R)={L\over\Re^2}{b^{2n}\over 2\pi n\Gamma(2n)}e^{-b\eta^{1/n}}~,
\end{equation}

\noindent where $\Gamma$ represents the complete Gamma function.
The photometric profile is thus characterized by two scales (e.g.,
$L$ and $\Re$) and one dimensionless parameter ($n$).  The
standard de Vaucouleurs (1948) profile is obtained by setting $n =
4$.

\subsection{Two--component $\dev$+exponential models}

For the surface brightness profile we take here the superposition
of an $R^{1/4}$ component and a standard ``exponential disk", i.e.

\begin{equation}
I(R) = I_{1/4}(0)\exp[-b(4)(R/\Rdev)^{1/4}] + I_{\rm
exp}(0)\exp[-b(1) (R/\Rexp)],
\end{equation}

\noindent where $\Rdev$ and $\Rexp$ are the effective radii of the
$\dev$ and of the exponential component, respectively. By fitting
this model to an observed photometric profile, we can derive the
total luminosities $\Lexp$, $\Ldev$ associated with the separate
components and from these the total luminosity $\Ltot=\Ldev
+\Lexp$. In conclusion, this two-component photometric model is
characterized by two scales ($\Ltot$ and $\Re$; for the
latter quantity we refer to the effective radius for the combined
profile) and by two dimensionless parameters (e.g., $\Lexp/\Ltot$ 
and $\Rexp/\Rdev$).

\subsection{$\finf$ models}

The $\finf$ models are {\it dynamical} models applicable to
collisionless stellar systems. They have been constructed (Bertin
\& Stiavelli 1984) by following the physical scenario that
elliptical galaxies may have formed through collisionless collapse
(van Albada 1982). In the spherical limit, their anisotropic
distribution function is given by

\begin{equation}
\finf = \cases{A(-E)^{3/2}\exp(-aE - cJ^2/2) &if $E\le0$,\cr 0 &if
$E>0$,\cr}
\end{equation}

\noindent where $E=v^2/2+\Phi(r)$ and $J$ are the star energy and
the magnitude of the star angular momentum, per unit mass,
respectively. Two of the three available dimensional constants
($A$, $a$, and $c$) set the physical scales of the model (e.g.,
the total mass and the half--mass radius $r_M$). The remaining
dimensionless parameter, for example written as
$\gamma=ac/(4{\pi}GA)$, then characterizes the one--parameter
family of $\finf$ models. The solution of the Poisson equation
provides the relation $\gamma = \gamma (\Psi)$, where $\Psi$
denotes the dimensionless central potential $\Psi=-a\Phi(0)$. In
practice, under the assumption that the stellar mass--to--light ratio is
constant within the galaxy and after projection for comparison
with the observations, the relevant photometric {\it profile}
associated with the $\finf$ models, in addition to a luminosity
scale ($L$) and a length scale ($\Re \approx r_M/1.3$), depends on
one dimensionless parameter only ($\Psi$). A detailed discussion
of the properties of this equilibrium sequence is given by Bertin
\& Stiavelli (1993). Here we limit our discussion to $\finf$
models characterized by $2\leq\Psi\leq 10$. In fact, models with
$\Psi<2$ are unstable, while, on the other hand, models with
$\Psi\geq 10$ are very similar to each other on the large scale,
except for the development of a nuclear concentration on scales
much smaller than those studied in this paper.

Given the focus of the present paper on the $R^{1/n}$ photometric
models, as a preliminary investigation we have compared the
photometric profiles of the $\finf$ dynamical models (treated as
``the galaxies" ) to those of the $\ser$ models (taken as the
``fitted models"). We have thus obtained the best-fit $n(\Psi)$ by
minimizing the quantity

\begin{equation}
\chi^2=\sum_{\eta_{\min}}^{\eta_{\max}}{[\mu_{\infty}(\eta)-\mu_{1/n}
(\eta)]^2},
\end{equation}

\noindent where $\eta = R/\Re$ is expressed in units of the
effective radius of the $\finf$ models. The adopted radial
interval is $0.1\leq \eta \leq 10$; note that the adopted outer
radius is well far out with respect to the radius $R_{\rm max}$
typically considered for observed photometric profiles.

%--------------------------------------------------------------------
\begin{figure}[htbp]
\parbox{1cm}{
\psfig{file=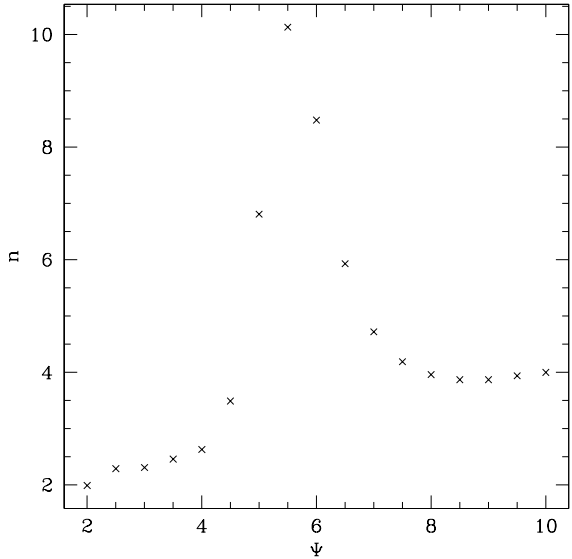,width=8cm,height=8cm,angle=0}} \caption[]{The
best-fit $n$, obtained by fitting the $\finf$ models, projected
along the line of sight, with $\ser$ profiles, as a function of
the model dimensionless central potential $\Psi$. The radial
interval adopted in the fit is described in the text. Note the
``plateau" at $n = 4$ reached by high-$\Psi$ models.}
\end{figure}
%-------------------------------------------------------------------

We found that for $\finf$ models with $\Psi~\gsim~5$ a fit over
the adopted radial interval can be obtained with $\Dmumax < 0.4$.
In contrast, for $\Psi~\lsim~ 5$, in order to obtain a fit with
$\Dmumax < 0.4$ we had to reduce the radial interval to typical
values $0.23\leq \eta \leq 4$. This smaller radial range is
actually applicable to realistic conditions. In these latter cases
of low-$\Psi$ models we found that the best-fit value of $n$ is
sensitive to the specific radial interval adopted; for example,
the quality condition $\Dmumax < 0.4$ can be satisfied for a
different radial range (say $0.1\leq \eta \leq 2$), but the
resulting best-fit value of $n$ turns out to be significantly
lower. In Fig. A.1 we show the optimal fitting relation
$n=n(\Psi)$ for the fits based on the full radial range, for $\Psi
\geq 5.5$, and on the reduced radial range ($0.23\leq \eta \leq
4$) for $\Psi < 5.5$. Note how close to the de Vaucouleurs
profiles, $n = 4$, are the high-$\Psi$ models, as realized from
the very beginning (see Bertin \& Stiavelli 1993).

\section{Fitting procedures}

In this Appendix we briefly describe different options to fit a
model, chosen within the set of models presented in Appendix A, to
the photometry of observed objects available within a given
spatial range.

\subsection{Profiles, circularized profiles, curves of growth}

In the procedure of {\it profile fitting}, we generally refer to
{\it circularized profiles}. The objects selected for the present
study are fairly round in projection. Therefore, the differences
between the photometric profiles taken along the short (with
coordinate $R_s$) or the long (with coordinate $R_l$) axis (as
preferred by CCD93, DCC94) with respect to the profiles obtained
by referring to the circularized radius $R = \sqrt{R_s R_l}$ are
rather small.

The observed profiles in $\magarsecs$, $\muobs(R)=-2.5\log\Iobs
(R)$, are compared to the corresponding model profiles $\Imod$

\begin{equation}
\mumod (R)=-2.5\log\Imod(R).
\end{equation}

\noindent Following standard methods (e.g., see Bertin, Saglia, \&
Stiavelli 1988), the best-fit parameters are determined by
minimizing the $\chi^2$ function defined as:

\begin{equation}
\chi ^2=\sum_{i=1}^N \frac{[\muobs (R_i)-\mumod (R_i)]^2}
{\sigobs^2 (R_i)(N-\Npar)},
\end{equation}

\noindent where $N$ is the number of data-points in the interval
$(\Rmin,\Rmax)$, $\sigobs (R_i)$ represents the photometric error
on $\muobs (R_i)$, and the factor $(N-\Npar)$ takes into account
the number of free parameters in the fit. The formal errors
$\delta_i$ associated with the best-fit values of the parameters
$x_i$ involved can be calculated from the covariance matrix ${\bf
E}=2{\bf H^{-1}}$, where ${\bf H}$ is defined as

\begin{equation}
H_{ij}=(N-\Npar)\frac{\partial ^2\chi ^2}{\partial x_i\partial
x_j}~.
\end{equation}

\noindent Therefore, we have

\begin{equation}
\delta_i = \sqrt{E_{ii}}~,
\end{equation}

\noindent while the off-diagonal terms provide the correlation
coefficients.

The parameters thus determined (e.g., the effective radius $\Re$)
do not represent the quantities implied by their ideal definitions
(e.g., the projected radius of the disk that encloses half of the
total luminosity), because the fit is performed only on a limited
radial range ($R_{\rm min}$, $R_{\rm max}$). In this procedure the
model parameters (e.g., $n$, $\Re$, and $I_0$ for the $\ser$
models) are taken to be independent of each other.

A second fitting procedure (preferred, e.g., by PS97) refers to
the so-called {\it curve of growth}, i.e. to the function
associated with the integrated luminosity $L(R) = 2\pi \int_0^R
R^{\prime} I (R^{\prime})dR^{\prime}$. One then compares the data
points for $m_{\rm obs} (R) = -2.5\log\Lobs (R)$, expressed in
magnitudes, to the corresponding curve of growth $m_{\rm mod} (R)$
derived from the photometric model. It is clear that this
procedure involves a major difference with respect to the profile
fitting procedure, because the data points of the curves of growth
cannot be treated as independent of each other. If we proceed {\it
naively}, but {\it improperly}, by means of a $\chi ^2$ function
defined as in Eq.~(B.2), (but in the results shown in Section 2.3 we took 
a constant photometric error) with $\mu$ replaced by $m$, we find
``best-fit values" of the relevant parameters that can be
significantly discrepant with respect to those found from profile
fitting (compare the values of $n$ given in the second column of
Table 3 with those recorded in Table 1). The discrepancies follow
the general trends reported in the literature (see Introduction
and PS97).

\subsection{Constraints}

One of the tests that we have performed on our set of observed
photometries has been the study of the dependence of the best-fit
parameters on the radial range of the fit. In order to compare
with earlier studies (e.g., see Burkert 1993), we have thus
considered a case where all photometric profiles for the four
galaxies studied in this paper were limited to a common excursion
of 4.5 mag (see second group of columns in Table 3).

Another natural possibility (see Bertin, Saglia, \& Stiavelli
1988) is to consider a fit under the constraint that the observed
profile and the model profile be characterized by the same
integrated luminosity within the radial range of the fit. To these
fits we refer as fits with constrained luminosity (see Sect. 2,
third group of columns in Table 3). Explicitly, we impose:

\begin{equation}
\int_{\Rmin}^{\Rmax}R\Iobs (R)dR= \int_{\Rmin}^{\Rmax}R\Imod
(R)dR~,
\end{equation}

\noindent so that the luminosity scale is set as a constraint and
the number of free parameters with respect to which the $\chi ^2$
function is minimized is reduced by one.

Similarly, for the fit to the curves of growth it is natural to
impose that $\Lobs (\Rmax)=\Lmod (\Rmax)$.

\subsection{Comments}

Not unexpectedly, when free parameters are available, such as $n$
or $\Psi$, the search for the best-fit model actually leads to
several possible minima of the relevant $\chi^2$ function. In
Table 2 we have recorded the properties (in parentheses) of a
secondary minimum available for NGC 1379, within the $f_{\infty}$
models, which appears to be particularly significant. In Tables 1
and 3 we have considered only the $\ser$ models corresponding to
the absolute minima found. For each galaxy, we have also checked
the properties of other (typically 4 or 5) models corresponding to
secondary minima, with $\chi^2$ within a factor of 3-4 with
respect to that associated with the absolute minimum. The related
model parameters turn out to be significantly different only when
the associated value of $\chi^2$ changes significantly. The values
of $n$ and $\Re$ thus identified tend to increase, or decrease,
together, but the photometric parameter $\phiFP$ remains
robustly stable.

The case illustrated briefly in the second group of columns in
Table 3, that of the reduced radial range, presents, not
surprisingly, the most complex behavior. Here we can even witness
the occurrence of ``crossing" of solutions. An example of this
phenomenon is provided by the galaxy NGC 4374. For this object the
absolute minimum over the reduced radial range is obtained for $n
\approx 14.5$ and $\Re\approx 1092$, with $\chi^2 \approx 0.34$
(see Table 3). A very near (when considering the $\chi^2$ value)
secondary minimum is $n \approx 9.9$ and $\Re\approx 302$, with
$\chi^2 \approx 0.55$. When fitting the profile of the same galaxy
over the extended radial range the absolute minimum is found (see
Table 1) at $n \approx 9.5$ with $\Re\approx 259$ and $\chi^2
\approx 0.46$. Thus, the solution corresponding to the absolute
minimum obtained over the the whole available radial range can
turn out to be associated with a secondary minimum when the fit is
extended over a reduced radial range.

\section{General properties of the velocity dispersion profiles of
simple dynamical models}

If we wish to draw consequences, from the observed trends in the
photometric profiles, on the structure of the FP, we have to take
into account some dynamical information. In practice, for the
objects under consideration we have to estimate the relevant
virial coefficient (see Appendix D). Since the dynamical quantity
that appears in the FP is the central (projected) velocity
dispersion (based on a suitable small aperture), it is clear that
our consequences will depend mostly on the assumed behavior, for
the velocity dispersion profiles, in the vicinity of the galactic
center. In this Appendix we will show that if we start from a
given density profile and then proceed to infer, as commonly done
(through the Jeans equations under simple assumptions on the
pressure tensor; e.g., see PS97), the velocity dispersion profile,
the resulting kinematical profile is {\it very sensitive} to the
specific limiting behavior (at very small radii) of the density
profile we consider. In many cases the implied kinematical profile
may be completely unrealistic (e.g., by developing a major
``central hole", not observed; see Bailey \& MacDonald 1981). This
sensitivity is unfortunate, because it makes it necessary to
address in detail the dynamical basis of the photometric models
used to fit the observations. In this respect, we should recall
here that the velocity dispersion profiles of the $f_{\infty}$
models (see Appendix A.3) are (generally realistic) monotonic,
decreasing with radius.

In the following we focus on spherically symmetric, isotropic,
one--component galaxy models with finite mass and, for simplicity,
limit our discussion to the {\it unprojected} velocity dispersion
only.

Suppose that the model density profile at small radii is
characterized by the asymptotic behavior

\begin{equation}
\rho (r)\sim r^{-\alpha},
\end{equation}

\noindent with $0\leq\alpha < 3$.  From the explicit solution of
the appropriate hydrostatic equilibrium condition,

\begin{equation}
\frac{d}{dr}[\rho(r)\sigr^2 (r)] = -G{\rho(r)M(r)\over r^2}~,
\end{equation}

\noindent it is readily shown that for $\alpha =0$ the velocity
dispersion reaches a finite (non--vanishing) value at the center
(dependent on the shape of the entire density profile). For
$0<\alpha <1$ we find

\begin{equation}
\sigr^2 (r)\sim r^{\alpha},
\end{equation}

\noindent and so $\sigr^2 (0)=0$ and a ``central hole" is present
(note that all $\ser$ models fall in this case, because close to
the center their density profile is characterized by $\alpha = 1 -
1/n$). When $\alpha=1$, we have

\begin{equation}
\sigr^2 (r)\sim -r\ln (r)~,
\end{equation}

\noindent and so the central value of the velocity dispersion
vanishes. Finally, for $1<\alpha <3$ the relation is

\begin{equation}
\sigr^2 (r)\sim r^{2-\alpha}.
\end{equation}

\noindent Thus, for $1<\alpha<2$ the central velocity dispersion
vanishes and a hole occurs; when $\alpha=2$ the central velocity
dispersion is finite (its value depends on the shape of the entire
density profile, as in the case $\alpha = 0$), and for
$2<\alpha<3$ it diverges.

In order to obtain an estimate of the velocity dispersion gradient
close to the center, we can rewrite Eq.~(C2) as

\begin{equation}
{d\sigr^2\over dr}=-G{M(r)\over r^2}
-G\frac{1}{\rho^2}\frac{d\rho}{dr}\int_r^{\infty}{\rho(x)M(x)\over
x^2}dx~.
\end{equation}

\noindent Let us now focus on the case $\alpha=0$, and thus
express the density profile as $\rho(r)\sim \rho(0) - \rho_1
r^{\beta}$. Then, it can be shown that for $0<\beta <2$ there is
indeed a hole in the velocity dispersion; in turn, for $\beta =2$
the direction of the slope of the velocity dispersion depends on
the density profile of the entire model, while for $\beta >2$ the
velocity dispersion is monotonically decreasing with radius.

We may add that, independently of the slope of the stellar density
profile in the central regions, if a massive black hole is
present, then the slope of the square of the velocity dispersion
is -1, and so, at least in the central regions, no hole is present
in the velocity dispersion profile.

By experimenting with more general density profiles in the central
regions, we have also encountered the possibility of ``wavy"
velocity dispersion profiles.

\section{The virial coefficient $\Kvir$}

After this short kinematical digression, it should be clear that
the virial coefficient $\Kvir = GM/(\Re\sigz^2)$ that controls the
relation between the virial theorem and the FP is sensitive to the
{\it dynamical model} that we adopt to describe galaxies, not just
to the photometric signature as captured, for example, by the
$\ser$ profiles. In addition, within a given set of dynamical
models, the virial coefficient can be sensitive to the precise
definition that we adopt for ``central velocity dispersion".

To set the basis for our simulations of Sect.~3, we have computed
the virial coefficient for the isotropic $\ser$ models, with
$\sigz$ based on an aperture of radius $\Re/10$. In addition, to
illustrate the above-mentioned sensitivity problem we have
computed the same coefficient also for the $\finf$ models. We have
checked that, for the isotropic $\ser$ models, the numbers are in
agreement with those reported earlier by PS97. The results are
summarized in Fig.~D.1.

%--------------------------------------------------------------------
\begin{figure}[htbp]
\parbox{1cm}{
\psfig{file=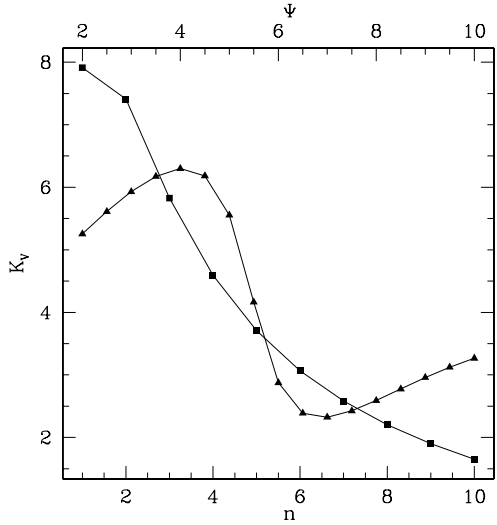,width=8cm,height=8cm,angle=0}}
\caption[]{The virial coefficient for the $\finf$ (triangles) and
the isotropic $\ser$ (squares) models based on an aperture of
radius $\Re /8$.}
\end{figure}
%-------------------------------------------------------------------

From this figure, an immediate warning in relation to the sensitivity
problem derives from inspection of the high-$\Psi$ models: while with
respect to the photometric performance the models with $7 \leq \Psi
\leq 10$ are all very well fitted by $\dev$ profiles (see Fig.~A.1),
their virial coefficient is found to vary significantly, from $\approx
2.3$ to $\approx 3.3$.

Another simple way to point out quantitatively the embarassing
modeling problem that we are facing is given in Table D.1. Here we
record, for the four galaxies studied in this paper, the value of
the virial coefficient obtained by referring to the best-fit
$\ser$ models suggested by Table 1, to those of Table 3 (b:
reduced radial range), or to the best-fit $\finf$ models
identified in Table 2. Note that if we modeled these galaxies as
isotropic $\dev$ models, they would all be assigned the same
virial coefficient $K_{1/4} \approx 4.6$. We should emphasize
that, in the spirit of the studies of the FP, all these inferences
are made {\it without} considering the detailed behavior of the
actual velocity dispersion profiles for the galaxies under
investigation.

%___________________________________
\begin{table}[hbt!]
\caption[]{Virial coefficient $\Kvir$ corresponding to different
fitting procedures.}
\begin{center}
\begin{tabular}{cccc}
\hline NGC  & $(\ser)^{\mathrm{a}}$   & $(\ser)^{\mathrm{b}}$
& $\finf$\\ \hline\hline 1379 & 6.8   & 6.8   & 3.2    \\
     &       &       & (5.1)  \\
4374 & 1.8   & 1.8   & 2.4    \\ 4458 & 5.7   & 5.5   & 3.0    \\
4552 & 1.6   & 1.8   & 2.5    \\ \hline
\end{tabular}
\end{center}
\begin{list}{}{}
\item[$^{\mathrm{a}}$] Extended radial range, no $L$ constraint. See
Table 1.
\item[$^{\mathrm{b}}$] Reduced radial range, no $L$ constraint. See
Table 3.
\end{list}
\label{Table D.1}
\end{table}
%___________________________________


\begin{thebibliography}{}

   \bibitem[1998] {} Andredakis, Y.C. 1998,
      MNRAS, 295, 725

   \bibitem[1995] {} Andredakis, Y.C., Peletier, R.F., \& Balcells, M. 1995,
      MNRAS, 275, 874

   \bibitem[1981] {} Bailey, M.E., \& MacDonald, J. 1981,
      MNRAS, 194, 195

   \bibitem[1992] {} Bender, R., Burstein, D., \& Faber, S.M. 1992,
      ApJ, 399, 462

   \bibitem[1993] {} Bender, R., Burstein, D., \& Faber, S.M. 1993,
      ApJ, 411, 153

   \bibitem[1998] {} Bender, R., Saglia, R.P., Ziegler, B., et al. 1998,
      ApJ, 493, 529

   \bibitem[1994] {} Bertin, G., Bertola, F., Buson, L.M., et al. 1994,
      A\&A, 292, 381

   \bibitem[1988] {} Bertin, G., Saglia, R.P., \& Stiavelli, M. 1988,
      ApJ, 330, 78

   \bibitem[1984] {} Bertin, G., \& Stiavelli, M. 1984,
      A\&A, 137, 26

   \bibitem[1993] {} Bertin, G., \& Stiavelli, M. 1993,
      Rep. Prog. Phys., 56, 493

   \bibitem[1992] {} Bower, R.G., Lucey, J.R., \& Ellis, R.S. 1992,
      MNRAS, 254, 601

   \bibitem[1993] {} Burkert, A. 1993,
      A\&A, 278, 23

   \bibitem[1993] {} Caon, N., Capaccioli, M., \& D'Onofrio, M. 1993,
      MNRAS, 265, 1013 (CCD93)

   \bibitem[1994] {} Caon, N., Capaccioli, M., \& D'Onofrio, M. 1994,
      A\&A, 106, S199

   \bibitem[1990] {} Caon, N., Capaccioli, M., \& Rampazzo, R. 1990,
      A\&A, 86, S429

   \bibitem[1987] {} Capaccioli, M. 1987,
      in: Structure and dynamics of elliptical galaxies, ed. P.T.
      de Zeeuw, Reidel, Dordrecht, p. 47

   \bibitem[1989] {} Capaccioli, M. 1989,
      in: The world of galaxies,
      eds. H.G. Corwin, L. Bottinelli,
      Springer-Verlag, Berlin, p. 208

   \bibitem[1985] {} Capelato, H.V., de Carvalho, R.R., \& Carlberg, R.G. 1995,
       ApJ, 451, 525

   \bibitem[1991] {} Ciotti, L. 1991,
      A\&A, 249, 99

%   \bibitem[1997] {} Ciotti, L. 1997,
%      in: Global scaling relations: origins, evolution, and
%      applications, eds. L.N. da Costa, A. Renzini,
%      Springer-Verlag, Berlin, p. 38

   \bibitem[1999] {} Ciotti, L., \& Bertin, G. 1999, A\&A, 352, 447

   \bibitem[1997] {} Ciotti, L., \& Lanzoni, B. 1997,
      A\&A, 321, 724

   \bibitem[1996] {} Ciotti, L., Lanzoni, B., \& Renzini, A. 1996,
      MNRAS, 282, 1

   \bibitem[1992] {} Ciotti, L., \& Pellegrini, S. 1992,
      MNRAS, 255, 561

   \bibitem[1996] {} Courteau, S., de Jong, R.S., \& Broeils, A.H. 1996,
      ApJ, 457, L73

   \bibitem[1988] {} Davies, I.G., Phillips, S., Cawson, M.G.M.,  et al. 1988,
      MNRAS, 232, 239

   \bibitem[1983] {} Davies, R.L., Efstathiou, G., Fall, S.M., et al. 1983,
      ApJ, 266, 41

   \bibitem[1988] {} de Carvalho, R.R., \& da Costa, L.N. 1988,
      ApJ, 68, S173

   \bibitem[1948] {} de Vaucouleurs, G. 1948,
      Ann. d'Astroph., 11, 247

   \bibitem[1979] {} de Vaucouleurs, G., \& Capaccioli, M. 1979,
      ApJ, 40, S699

   \bibitem[1991] {} de Vaucouleurs, G., de Vaucouleurs, A., Corwin, H.G., 
      et al. 1991,
      Third Reference Catalogue of Bright Galaxies,
      Springer-Verlag, New York

   \bibitem[1995] {} Djorgovski, S. 1995,
      ApJ, 438, L29

   \bibitem[1987] {} Djorgovski, S., \& Davis, M. 1987,
      ApJ, 313, 59

   \bibitem[1994] {} D'Onofrio, M., Capaccioli, M., \& Caon, N. 1994,
      MNRAS, 271, 523 (DCC94)

   \bibitem[1987] {} Dressler, A., Lynden-Bell, D., Burstein D.,
      et al. 1987,
      ApJ, 313, 42

   \bibitem[1987] {} Faber, S.M., Dressler, A., Davies, R.L., et al. 1987,
      in: Nearly normal galaxies, ed. S.M. Faber, Springer, New York,
      p. 175

   \bibitem[1997] {} Gerbal, D., Lima-Neto, G.B., Marquez, I., \& Veraghen, 
      H. 1997,
      MNRAS, 285, L41

   \bibitem[2001] {} Gerhard, O., Kronawitter, A., Saglia, R.P., \& Bender, 
      R. 2001,
      AJ, 121, 1936

   \bibitem[1998] {} Graham, A.W. 1998,
      MNRAS, 295, 933

   \bibitem[1996]{} Graham, A.W., Lauer, T.R., Colless, M., \&
      Postman, M.  1996, ApJ, 465, 534

   \bibitem[1998]{} Graham, A.W., Colless, M., Busarello, G., Zaggia, S., 
      \& Longo, G. 
      1998,
      A\&AS, 133, 325

   \bibitem[1997] {} Graham, A.W., \& Colless, M. 1997,
      MNRAS, 287, 221

   \bibitem[1995] {} Hjorth, J., \& Madsen, J. 1995,
      ApJ, 445, 55

   \bibitem[1993] {} J{\o}rgensen, I., Franx, M., \& Kj{\ae}rgaard, P.
      1993,
      ApJ, 411, 34

   \bibitem[1999] {} J{\o}rgensen, I., Franx, M., Hjorth, J., et al.
      1999,
      MNRAS, 308, 833

   \bibitem[1997] {} Kelson, D.D., van Dokkum, P., Franx, M., et al.
      1997,
      ApJ, 478, L13

   \bibitem[2000] {} Kelson, D.D., Illingworth, G.D., van Dokkum, P.,
      et al. 2000a,
      ApJ, 531, 137

   \bibitem[2000] {} Kelson, D.D., Illingworth, G.D., van Dokkum, P.,
      et al. 2000b,
      ApJ, 531, 159

   \bibitem[2000] {} Kelson, D.D., Illingworth, G.D., van Dokkum, P.,
      et al. 2000c,
      ApJ, 531, 184

   \bibitem[2000] {} Khosroshahi, H.G., Wadadekar, Y., Kembhavi,
      A., \& Mobasher, B. 2000,
      ApJ, 531, L103

   \bibitem[1995]{} McElroy, D.B. 1995,
      ApJS, 100, 105

   \bibitem[1990]{} Makino, J., Akiyama, K., \& Sugimoto, D. 1990,
      PASJ, 42, 205

   \bibitem[1985] {} Michard, R. 1985,
      A\&A, 59, S205

   \bibitem[1997] {} Pahre, M.A., \& Djorgovski, S.G. 1997,
      in: The Nature of Elliptical Galaxies, eds. M. Arnaboldi, G.S. Da Costa,
      \& P. Saha, ASP Conf. Ser. vol. 116, p. 154

   \bibitem[1998] {} Pahre, M.A., de Carvalho, R.R., \& Djorgovski, S.G.
      1998,
      AJ, 116, 1606

   \bibitem[1998] {} Pahre, M.A., Djorgovski, S.G., \& de Carvalho, R.R.
      1998,
      AJ, 116, 1591

   \bibitem[1997] {} Prugniel, P., \& Simien, F. 1997,
      A\&A, 321, 111 (PS97)

   \bibitem[1993] {} Renzini, A., \& Ciotti, L. 1993,
      ApJ, 416, L49

   \bibitem[1997] {} Saglia, R.P., Bertschinger, E., Baggley, G.,
      et al. 1997,
      ApJ, 109, S79

   \bibitem[1986] {} Schombert, J.M. 1986,
      ApJ, 60, S603

   \bibitem[1995] {} Scorza, C., \& Bender, R. 1995,
      A\&A, 293, 20

   \bibitem[1968] {} Sersic, J.L. 1968,
      Atlas de galaxias australes.
      Observatorio Astronomico, Cordoba

   \bibitem[2001] {} Stiavelli, M., Miller, B.W., Ferguson, H.C.,
      et al. 2001,
      AJ, 121, 1385

   \bibitem[1999] {} Treu, T., Stiavelli, M., Casertano, S., et al.
      1999,
      MNRAS, 308, 1037

   \bibitem[2001] {} Treu, T., Stiavelli, M., M{\o}ller, P., et al.
      2001a,
      MNRAS, 326, 221

   \bibitem[2001] {} Treu, T., Stiavelli, M., Bertin, G., et al.
      2001b,
      MNRAS, 326, 237

   \bibitem[2001] {} Trujillo, I., Graham, A.W., \& Caon, N. 2001,
      MNRAS, 326, 869

   \bibitem[1982] {} van Albada, T.S. 1982,
      MNRAS, 201, 939

   \bibitem[1995] {} van Albada, T.S., Bertin, G., \& Stiavelli, M. 1995,
      MNRAS, 276, 1255

   \bibitem[1996] {} van Dokkum, P., \& Franx, M. 1996,
      MNRAS, 281, 985

   \bibitem[1998] {} van Dokkum, P., Franx, M, Kelson, D.D., et al. 1998a,
      ApJ, 500, 714

   \bibitem[1998] {} van Dokkum, P., Franx, M, Kelson, D.D., et al. 1998b,
      ApJ, 504, L17

   \bibitem[1999] {} Wadadekar, Y., Robbason, B., \& Kembhavi, A. 1999,
      AJ, 117, 1219

   \bibitem[1994] {} Young, C.K., \& Currie, M.J. 1994,
      MNRAS, 268, L11

\end{thebibliography}
\end{document}